# The Absolutism of Space and Time with Variable Scalars and Gravitational Theory as well as Cosmology Established in Flat Reference Frame

(1)

**Mei   Xiaochun**

(Department of Physics, Fuzhou University, China, E-mail: mxc001@163.com)

## Content

  **Authors who are not interested in the concept analysis of space-time and gravitation can start from Section 4 after reading introduction**

# Introduction

The Einstein's theories of space-time and gravity as well the stander cosmology are reconstructed thoroughly in this paper based on flat reference frame. The rational parts of the Einstein's theories are reserved while the irrational parts including space-time paradoxes and singularities and so on are eliminated completely. A more rational the theory of space-time, gravitation and cosmology is established.

By transforming the geodesic equation described by the Schwarzschild solution of the Einstein's equation of gravitational field into flat reference frame for description, the revised formulas of the Newtonian gravitation is obtained. Based on the formulas, all experiments which support general relativity can also be explained well, but the theory has no any space-time singularity and other strange characters which exist in general relativity. It is pointed that there exists logical difficulties to use the equivalent principle to explain spectrum gravitational redshift actually. The new formula of gravitational redshift and the revised Doppler formula of redshift when light's source moves in gravitational field are deduced. Based on the Schwarzschild solution of the Einstein's equation of gravitational field, it is proved that the speed of light would change and the isotropy of light's speed would be violated in gravitational field with spherical symmetry. It is suggested to use the method of the Michelson—Morley interference to verify the change of light's speed and the violation of isotropy in the gravitational field of the earth. This experiment can be considered as a new verification for general relativity in the weak gravitational field with spherical symmetry. The theory may also be used to explain so-called the Pioneer Anomaly which the Einstein's theory of gravitation can not do.

It is proved that according to the Robertson---Walker metric, the velocity of light which is emitted by the celestial bodies in the expansive universe would obey the Galileo's addition ruler of classical mechanics, in stead of the Einstein's addition ruler in special relativity. This result violates the principle of invariance of light's velocity and contradicts with the current physical experiments and astronomical observations. So the Robertson---Walker metric can not be used as the basic frame of space-time to describe the expensive universe. The theoretical foundation of modern cosmology must be reestablished. Some established viewpoints and conclusions should be re-surveyed. A most direct result is that the affirmation about dark energy governing our universe and the universal accelerating expansion deduced from the fit between the cosmological theory and the observations of high red-shift Ia type supernovae are untrue. When the revised Newtonian formula of gravitation is used to discuss the problem of the universal expansion, the new motion equation of cosmology can be obtained, which is similar to the Friedmann equation of the current cosmology but does not contain cosmic constant. By means of it, the Hubble diagrams of high red-shift Ia type supernovae can be explained well. The hypothesis of dark energy becomes unnecessary and the problem of cosmic constant which has puzzled physical circle for a long tome can be get rid of thoroughly. Meanwhile, we do not need to suppose that more than three fourth of total material in our universe would be non-baryon dark material, if they exist indeed. The problem of the universal age can also be solved well.

Similar to electromagnetic theory, by introducing magnetic-like gravitation, a more rational theory of gravitation can be established in flat reference frame without any singularity and morbid character. The theory may be consistent with quantum mechanics and may be renormalizable after being quantized. So it may provide a really reliable foundation for the unity of four interaction forces.

Since the Galileo's time, the velocity of object's motion had been considered as a relative concept. However, by means of the dynamical effects of special relativity, it is proved that the velocity caused by accelerating process should be an absolute quantity, instead of a relative quantity. But this observable effect does not exist in the Newtonian mechanics, only by considering the dynamic effect of special relativity, it can be found. Therefore, the traditional idea that the velocity is a relative concept should be given up. The influence of accelerating process should be considered in space-time theory. Besides the Newtonian absolute theory of space-time with invariable scales and the Einstein's relative space-time theory with variable scales, there exists the third space-time theory, i.e., the absolute space-time theory with variable scales. This kind of theory of space-time, satisfying the demand of the modern cosmology, is more rational and more consistent with practice.

It is pointed that we only prove that inertial static mass and gravitational static mass are equivalent by the *Eötvös* type of experiments. But this type of experiments can not prove whether or not the inertial moving mass is also equivalent to gravitational moving mass. By means of the dynamic effect of special relativity, it can also be proved that the inertial moving mass is not equivalent to gravitational moving mass, or gravitation and inertial force are not equivalent locally. In the same, this kind of non-equivalence does not exist in the Newtonian mechanics. It is just Einstein himself who neglected the influence of dynamic effects of special relativity when he put forward the principles of space-time relativity and equivalence. By transforming the geodesic equation described by the Schwarzschild solution into flat space-time to describe, we can obtain the concrete form of gravitational moving mass automatically. Meanwhile, by means of abundant facts and logical analysis, the principle of general relativity is completely untenable.

By the coordinate transformations of the Kerr and Kerr-Newman solutions, the solutions for the static distributions of mass thin loop and double spheres with axial symmetry are obtained. The results indicate that no matter what the masses and the densities of thin loop and double spheres are, the space's curvatures in the centre points of loop and the double sphere's connecting line are infinite. The singularity points are completely exposed in vacuum. The space curvatures nearby the singularity points and the surfaces of thin loop as well as double spheres are also highly curved even though the gravitational fields are very weak. It is obvious that all of them are actually impossible. The results show that the space-time singularities appearing in the Einstein's theory of gravity are not caused by the high density and huge mass's distributions. They are caused actually by the descriptive method of curved space-time, having nothing to do with the real world. After the geodesic equation described by the Schwarzschild solution of the Einstein's equation of gravitational field is transformed into flat reference frame to describe, all space-time singularities are transformed into the infinite points of gravitations. This kind of infinites would appear in all theories based on the mode of point particles and not worth to be surprise. So we should give up the descriptive method of geometry, retuning to the classical descriptive method of dynamics and interaction for gravitation.

# Section 1  Absoluteness of Velocity Caused by Accelerating Process and the Absolutism of Space and Time with Variable Scalars

----The third logically consistent and more rational theory of space-time

## 1. Introduction

Since the Galileo's times, the velocity of object's motion had been considered as a relative concept. However, by means of the dynamic effect of special relativity, it is proved that the velocity caused by the accelerating process is not a relative quantity. It is an absolute quantity which can be determined by experiments actually. For the same reason, mass, length and time are also relative to accelerating process and with absolute significance. The influence of accelerating process should be considered in theory of space-time.  In logic, besides the Newtonian absolute theory of space-time with invariable scalars and the Einstein's relative theories with variable scalars, there exists the third absolute space-time theory with variable scalars. It is artificial and irrational to divide space-time theory into two parts---- kinematics and dynamics in special relativity. The real rational and logically consistent theory of space-time can not be based on the inertial reference frame. We should establish the theory of space-time on dynamic foundation universally. The concept of non-equivalent inertial reference frame is proposed in the paper. The real meaning of light's speed invariable is that we can not let the speed of an object with static mass surpass light's speed in vacuum. For the motion which is not caused through accelerating process, it is allowed to surpass light's speed in vacuum. For example, when the observers located on the rotating earth observe the distant stars, the tangent velocities of all stars greatly surpass light's velocity in vacuum. Based on the symmetrical consideration of space-time contractions, it can be proved that the absolutely stationary reference frame exists. In order to eliminate multifarious space-time paradoxes in special relativity and reach the consistence with the demand of the big-bang cosmology, the existence of the absolutely stationary reference frame is necessary. We can take the reference frame in which the microwave background radiation is isotropy as the absolutely resting reference frame. This kind of space-time theory is logically consistent without any internal contradiction, does not conflict with known physical experiments and astronomic observations, can be used to describe the real world well.

## 2. The possibility and necessity of the third space-time theory

As we known that according to the Newtonian theory, the scales of space-time were absolute and invariable, independent of the selection of reference frames. The absolutely resting reference frame was considered to be existent, though Newton did not know how to determine it. Reversely, according the Einstein's theory, the scales of space-time were relative and variable, dependent of the selection of reference frames. The absolutely resting reference frame was considered not to exist.

The relativity concept of motion had bee formed since the Galileo's age. In his two famous dialogues, Galileo demonstrated the phenomena that observers who are in a closed vessel could not find vessel's motion. This kind of relativity of velocity can keep the Newtonian law of mechanics unchanged. Einstein took the relativity concept of motion as one of the principles to establish his special relativity. But as shown below, by considering the dynamic effect of special relativity, the velocity caused by accelerating process can be proved to be an absolute quantity, in stead of relative quantity. This is completely different from the Newtonian and Einstein's theories. From this result, we can deduce the conclusion that space-time is with absolute significance and leads to the existence of absolutely resting reference frame. On the other hand, from the angle of logical possibility, the third space-time theory is allowed to exist besides the Newtonian and Einstein's theories. That is the absolute theory of space-time with variable space-time scales. The measurements of space-time are independent of the selection of reference frames. This result is completely different from that of the Newtonian and Einstein's theories. It would lead to the existence of the absolutely resting reference frame which can be determined by experiments.

Lots of experiments show that the scales of space-time are variable, so the Newtonian space-time theory is considered incorrect. Though the Einstein's theory has obtained great success, multifarious space-time paradoxes can be deduced from it[1]. But many physicists denied the existence of space-time paradoxes at present. They persist on that those paradoxes are caused only by incorrect understanding of relativity, and propose many methods to explain the paradoxes. However, it can be said that most of the explanations are ambiguous and even completely wrong. For example, in order to explain the falling paradox of sliding block (a kind of length paradox), the shape of sliding block was considered different in different inertial reference frames. In the resting reference frame, the shape of sliding board was regarded as to be flat, but when transformed into the moving inertial frame, the shape of sliding block was thought becoming parabola's shape[2]. This explanation is completely wrong and unacceptable. An object with flat shape in an inertial frame would still be the same shape in another inertial frame. Otherwise the principle of relativity would be violated. As for the most famous twin paradox with the existence of accelerating processes, general relativity or equivalent principle has to be used to explain it. This paradox is considered to be solved well at present[3]. However, the problem is not so simple. If we take more symmetrical form, this paradox is impossible to solve. For example, we send two twins into space by two rockets in opposite directions simultaneously. Then let them return to the earth at the same time after arbitrary period of time's travel. Because the situations are completely symmetrical for two twins, each twin would think that his brother is younger than himself if time contraction is relative. In this completely symmetrical case, general theory of relativity is also incapable. In fact, as we known, general relativity itself also demands certain absoluteness. The situation hints us that Einstein's completely relative space-time theory can not be true. It is unsuitable to use such theory to describe real world.

There are lots of discussions about the paradoxes existing in special relativity since the theory was put forward. This paper does not discuss them again. The key problem is how to obtain a new and more rational theory to substitute the Einstein's theory of relativity. This kind theory should reserve the excellence of Einstein's theory but avoid its disadvantage. This is real purpose for us to criticize the Einstein's theory of space-time. Because we lack this kind of theory at present, physicists prefer to bear the paradoxes of special relativity to give it up. As discussed below, the absoluteness concept of velocity caused by accelerating processes is the key for us to reach the correct theory of space-time. This kind of absolution can not be found in the Newtonian theory. Only by considering the dynamic effects of spherical relativity, it can be distinguished. The dramatic is that it was just Einstein himself who neglected this effect.

According to the common understanding at present, Einstein's space-time theory has nothing to do with the accelerating process. The theory only considers the measurement relation of space-time between two inertial reference frames. In special relativity, two reference frames are hypothesized to be at rest each other in beginning so that same space-time units (ruler and clock) can be defined. Then by introducing a relative speed between two them, we can deduce the Lorentz transformation formula based on the principles of relativity and invariability of light's speed. From the Lorentz transformation, we can reach a result that space-time's contractions are relative. However, we meet an inevitable problem here. That is, accelerating process is needed if we want to introduce a relative speed between two reference frames which are at rest in beginning. At least, one of two frames has to be accelerated. Thus, would accelerating process affect space-time characteristics? According to the current understanding, the effect of accelerating process can be neglected in space-time theory, for the Lorentz transformation is only relative to speed and speed is considered only to be a relative concept. However, in order to eliminate time paradox in special relativity, we have to consider the effect of accelerating or non-inertial process on space-time's nature. According to equivalent principle, inertial force is equivalent to gravitation, or non-inertial reference frame is locally equivalent to gravitational field. Because space-time scales would change in gravitational field, acceleration process would affect space-time's nature if equivalent principle holds. In this way, we meet a logical absurdity. On the one hand, in order to keep space-time relativity, we have to suppose that accelerating process does not affect space-time's nature. On the other hand, in order to reach a proper gravitational theory, we have to suppose that accelerating process is equivalent to at resting in gravitational field and gravitation field would affect space-time's nature. This kind of space-time theory is deviant. It tells us actually that a rational space-time theory can not avoid non-inertial motion's effects. Space-time's scales would be connected with accelerating process, speaking strictly, the action of both factors---- acceleration and accelerated time.

Einstein's special relativity is divided into two parts. One is kinematics independent of force and acceleration. Another is dynamics relative to force and acceleration. In kinematics, various paradoxes appear owing to velocity's relativity. In dynamics, there is no paradox because force and accelerating process are not relative. However, it is completely factitious and irrational to divided special relativity into such two parts. Non-inertial process's effect should be considered in rational space-time theory. We should also establish space-time theory consistently on the foundation of dynamics. Only after the dynamic theory of describing space-time's nature is established independently, we can conjecture the relation between non-inertial motion and gravitation. Otherwise the problems would entangle each other before non-inertial motion's influence on space-time has been known.

Therefore, we have to consider the third space-time theory with absoluteness and variable scales. In this theory, velocity caused by accelerating process can be proved to be an absolute concept and the absolutely resting frame is proved to be existent. This result also coincides with modern cosmology. We can take the inertial reference frame which was at rest together with the original point of Big-bang universe as the absolutely resting reference frame. All other reference frames in universe are considered to move relative to it.

### 3. The absoluteness and anisotropy of velocity caused by accelerating process

We now prove that velocity is an absolute concept by means of an ideal experiment. As shown in Fig. 1.1, suppose there are three resting reference frames in vacuum represented by $K_0$, $K_1$ and $K_2$ respectively. $K_1$ and $K_2$ are two closed chambers. There are two objects $A_1$ and $A_2$ with same masses in both chambers individually. Each object is connected to two sides of the chamber by two elastic ropes. The nature and length of elastic ropes are completely the same. Then $K_2$ is accelerated in a constant acceleration $a$ toward to right side, while $K_0$ and $K_1$ keep at rest. After a long enough time, the speed of $K_2$ reaches $V_2 = V'$. According to the dynamical theory of special relativity, the inertial mass of object $A_2$ and the inertial force acting on the elastic rope caused by object $A_2$ are individually

$$m_2 = \frac{m_0}{\sqrt{1 - V'^2/c^2}} > m_0 \qquad F_2 = \frac{d}{dt} \frac{m_0 V'}{\sqrt{1 - V'^2/c^2}} = \frac{m_0 a}{\left(1 - V'^2/c^2\right)^{3/2}} \qquad (1.1)$$

Therefore, the elastic rope on the right side of object $A_2$ would be pulled longer and longer in accelerating process. Suppose the biggest force that elastic rope can bear is $F_2$. When $V_2 > V'$, elastic rope would be pulled apart. Suppose that accelerating process stops when the speed of $K_2$ just reaches $V_2 = V'$. In this case, the elastic rope has not yet been pulled apart. After that, $K_2$ moves in a uniform speed $V'$ towards right side, and elastic rope resumes to original length again (though there exists the Lorentz contraction, it is unimportant in this discussion.) for there is no force acting on it again.

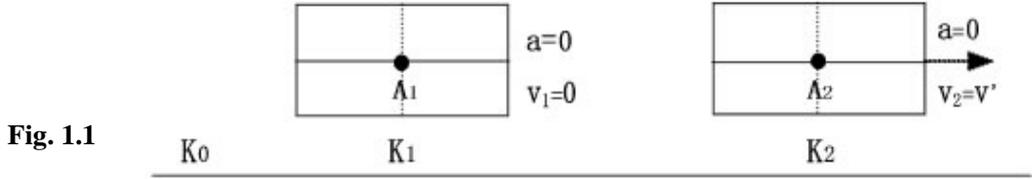

**Fig. 1.1**

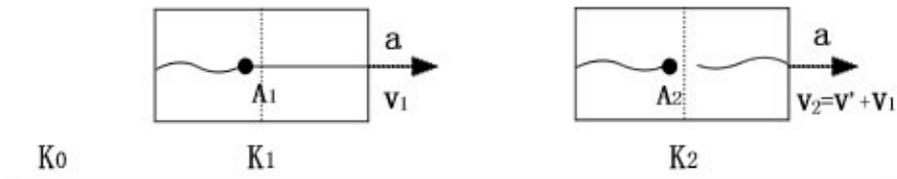

**Fig. 1.2**

Because $K_2$ is in inertial state again, according to space-time relativity, observers in $K_2$ would think that $K_2$ reference frame is at rest, while $K_0$ and $K_1$ reference frames move in uniform speed $V'$ towards left side. So according to observers in $K_2$, object $A_2$ is at rest with a mass $m_0$ and object $A_1$ is moving with a mass $m_1 = m_0 / \sqrt{1 - V'^2 / c^2} > m_0$. This opinion is just opposite to that of observers in $K_0$ and $K_1$. In the viewpoint of observers in $K_0$ and $K_1$, $K_1$ is at rest in all time and $K_2$ moves towards right side in uniform speed $V'$. So the mass of object $A_1$ is $m_0$ and the mass of object $A_2$ is $m_2 > m_0$. Therefore, mass becomes a relative concept for observers in two reference frames with a relative speed between them.

However, it can be proved that this relative idea is actually impossible as analyzed below. If we want to know an object's mass, we need to accelerate it. After that, we can decide its mass by the relation between acceleration and force acted on it. As shown in Fig.1.2, after $K_2$ reaches a uniform velocity $V'$, we accelerate simultaneously both $K_1$ and $K_2$ again relative to $K_0$ in the same acceleration $a$ towards right side for a short time. After $K_1$ reaches a small speed $V_1$ ($0 < V_1 < V'$), accelerating process is stopped and $K_1$ and $K_2$ return to the states of uniform motions again. Relative to observers in resting $K_0$, the speed of $K_2$ becomes $V_1 + V_2 > V'$, so that the elastic rope connected to the right side of object $A_2$ would be pulled apart immediately, for the inertial force acted on the rope is bigger than $F_2$. On the other hand, because $K_1$ is only accelerated for a short time and its speed is far small than $V'$, the elastic rope connected to the right side of object $A_1$ would not be pulled apart. Because whether the elastic ropes are pulled apart or not are absolute events, observers in $K_1$ and $K_2$ would observe the same phenomena and admit these facts.

Thus, we immediately see that the things happen in two inertial reference frames with a relative uniform speed between them are completely different. We can let observers in $K_1$ and $K_2$ are in the state of dormancy from beginning to the time that $K_2$ reaches a uniform speed $V'$, then awake them. In this case, the observers in two reference frames do not known which one has been accelerated. After that, we accelerate $K_1$ and $K_2$ for a while again. By means of the facts that the rope connected to object $A_2$ is pulled apart, but the rope connected to object $A_1$ is not pulled apart, observers in both $K_1$ and $K_2$ reference frames can obtain the same viewpoint that $K_2$ has been accelerated and has obtained a speed $V'$. Meanwhile, $K_1$ has always been at rest before the second accelerating process is carried out. There is no relativity in this case. Velocity caused by accelerating process becomes an absolute quantity. It is obvious that this result violates the principle of space-time relativity.

As we know that in the Newtonian theory, object's mass does not change with speed, so elastic rope would not be pulled apart in the process above. It means that we can not decide which reference frame moves and which is at rest by the same experiment according to the Newtonian theory. So in the Newtonian theory, velocity is still a relative concept. It is just by means of the dynamical effect of special relativity, similar to acceleration, the velocity produced by accelerating process becomes absolute.

Therefore, moving mass of an object should be an absolute quantity. When there exists a relative speed between $K_1$ and $K_2$, observers in both reference frames would have the same viewpoint about the moving masses of objects $A_1$ and $A_2$. If $K_2$ is accelerated, the mass of object $A_2$ increase. Because $K_1$ is at rest, the mass of object $A_1$ is unchanged. There is no relativity here. The accelerating process has an absolute effect on an object's mass. This is easy to understanding from the angle of energy conservation. Energy is needed when a force is acted on an object and work is done. The result is that object's energy or motion mass increases. Therefore, different from traditional idea, object's motion has reason. Some observable and measurable results would be caused when the object obtain a velocity through accelerating process. Unfortunately in the Newtonian and Einstein's theories, this point is neglected so that velocity becomes something that can be designated arbitrary without origin.

**Fig. 1.3** 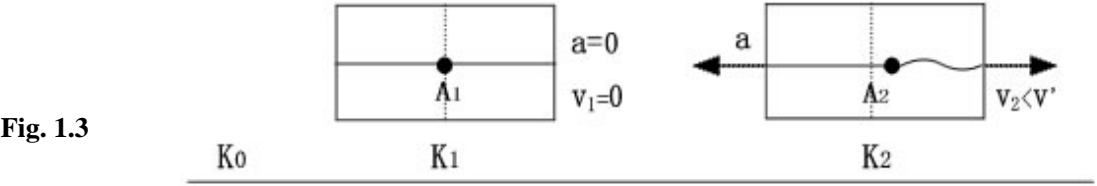

Now let us to show that velocity caused by accelerating process is anisotropic in general. For this purpose, let's discuss the decelerating process of $K_2$. As shown in Fig. 1.3, relative to resting $K_0$ and $K_1$, after $K_2$ has just reached a uniform speed $V'$ without rope being pulled apart, we accelerate $K_2$ towards left side in the same acceleration $a$ again (actually decrease $K_2$). In beginning, the velocity of $K_2$ is still towards to right side. After a period of time, the speed of $K_2$ is decreased to zero, so that three reference frames become at rest each other again. Because in the whole process the speed of $K_2$ is always small than $V'$, the rope connected to the left side of object $A_2$ would not be pulled apart. Because whether the rope is pulled apart or not is an absolute event, observers in $K_2$ would also observer the same result and hold the same viewpoint.

Thus we can see immediately that the results are completely different in both cases that $K_2$ are accelerated towards right side and towards left side. When it is accelerated towards right side, the rope would be pulled apart. But when it is accelerated towards left side, the rope would not be pulled apart. All of observers in three frames agree with this result. So observers in reference frame $K_2$ would think that space is with anisotropy for them. They can judge themselves moving along the direction at which velocity increases or along the direction at which velocity decreases by whether or not the ropes are pulled apart.

Obviously, this result also violates the principle of space-time relativity.

## 4. The nonequivalent inertial frames and the absoluteness of the Lorentz contraction

So we have to consider the effect of accelerating process on space-time's natures. At first, let's introduce the concepts of nonequivalent inertial reference frames and equivalent inertial reference frames. We call such two reference frames as nonequivalent inertial reference frames. In beginning, these two reference frames are at rest each other and are defined the same unit scales of space and time (ruler and clock). Then by accelerating one of them at least, a motion velocity is introduced between them. After accelerating process is stopped, they become inertial reference frames again with a relative velocity. That is to say, the relative velocity between two nonequivalent inertial frames is produced by accelerating process.

In fact, all common activities and scientific experiments of humankind are carried out on nonequivalent reference frames. The measurements about the relation of space and time are carried out in nonequivalent reference frames. For example, the Michelson interferometer's two arms are just nonequivalent reference frames when we use the interferometer to verify the invariability principle of light's speed. When one of arms moves along with the earth's moving direction, another arm can be considered to be at rest. When the interferometer rotates, the motion states of two arms exchange each other through non-inertial rotating. In fact, the earth is just a big non-inertial system, all scientific experiments done on it involves the effects of non-internal motion. Only based on nonequivalent reference frames, the theory is meaningful. Once theory is established on nonequivalent reference frames, absoluteness and asymmetry would be introduced.

Therefore, the two foundational hypothesis of Einstein's special relativity should be revised as:

1. Accelerating processes would affect the natures of space-time of nonequivalent inertial reference frames and the effects are absolute and measurable.

2. The speed of light is invariable among arbitrary non-inertial reference frames in vacuum.

By means of these two hypotheses, the same Lorentz transformation can also be deduced, but the transformation has absolute significance now. In order to obtain the transformations, we suppose that there are two reference frames $K_1$ and $K_2$ in vacuum. In beginning they are at rest each other and the same rulers and clocks are defined for them. Then let $K_1$ is still at rest and $K_2$ is accelerated. When the speed of $K_2$ reaches $V$, accelerating process is stopped. After that, $K_2$ moves in a uniform speed relative to $K_1$. Suppose at a certain moment, the coordinate point $x_1'$ located at $K_1$ meets the coordinate point $x_2'$ located at $K_2$. At this time, the clock at point $x_1'$ indicates time $t_1'$, and the clock at point $x_2'$ indicates time $t_2'$. Meanwhile, a bind of light is send out along the parallel $X$ axes of two frames. This light reaches the point $x_1$ at time $t_1$ relative to $K_1$, reaches the point $x_2$ at time $t_2$ relative to $K_2$. In light of the invariability principle of light's speed between non-inertial reference frames, we have

$$(x_1 - x_1')^2 - c^2(t_1 - t_1')^2 = 0 \qquad (x_2 - x_2')^2 - c^2(t_2 - t_2')^2 = 0 \qquad (1.2)$$

From these two equations, we can deduced the Lorentz formulas

$$x_2 - x_2' = \frac{x_1 - x_1' - V(t_1 - t_1')}{\sqrt{1 - V^2/c^2}} \qquad t_2 - t_2' = \frac{t_1 - t_1' - V(x_1 - x_1')/c^2}{\sqrt{1 - V^2/c^2}} \qquad (1.3)$$

Suppose space and time intervals for two reference frames are $\Delta l_1 = x_1 - x_1'$, $\Delta l_2 = x_2 - x_2'$, $\Delta t_1 = t_1 - t_1'$ and $\Delta t_2 = t_2 - t_2'$ individually  we can obtain the Lorentz contraction formulas

$$\Delta l_1 = \Delta l_2 \sqrt{1-\frac{V^2}{c^2}} \qquad\qquad \Delta t_1 = \frac{\Delta t_2}{\sqrt{1-V^2/c^2}} \qquad (1.4)$$

It shows that the length of rule fixed on $K_2$ becomes short and the time of clock fixed on $K_2$ becomes slow comparing with that fixed on $K_1$. Because the motion speed of $K_2$ is caused by accelerating process, we should think that the real change of space-time's scales of $K_2$ reference frame takes place owing to the effect of non-inertial process. Though the formulas of space-time contractions are relative to speed, it is only a superficies. The effects of accelerating process are hided. When accelerating process stops, space-time's scales are fixed.

On the other hand, we let $K_2$ at rest and accelerate $K_1$ towards left side in the beginning. When speed of $K_2$ reaches $-V$, stop accelerating process, we obtain the conversed Lorentz transformation

$$x_1 - x_1' = \frac{x_2 - x_2' + V(t_2 - t_2')}{\sqrt{1-V^2/c^2}} \qquad\qquad t_1 - t_1' = \frac{t_2 - t_2' + V(x_2 - x_2')/c^2}{\sqrt{1-V^2/c^2}} \qquad (1.5)$$

As well as the conversed space-time contraction formulas

$$\Delta l_2 = \Delta l_1 \sqrt{1-\frac{V^2}{c^2}} \qquad\qquad \Delta t_2 = \frac{\Delta t_1}{\sqrt{1-V^2/c^2}} \qquad (1.6)$$

It means that the length of ruler fixed on $K_1$ becomes short and the time of clock fixed on $K_1$ becomes slow comparing with that fixed on $K_2$. Because the motion speed of $K_1$ is caused by accelerating process, we should think that the real change of space-time's scales of $K_1$ reference frame takes place owing to the effect of non-inertial process. So it can be seen that the premises of the Lorentz transformation formulas (1.3) and (1.5) are different. Which reference frame is accelerated, of which space-time scales are changed and the changes are also real and absolute. In fact, before Einstein proposed the explanation of relativity, Lorentz thought that space-time's contraction was absolute effect owing to the motions of objects relative to static ether. We used to regard reference frames as some things which were abstract and arbitrary velocity could be endowed to. However, any practical reference frame is always composed of material with mass. Force is always needed when a practical reference frame is accelerated. The formulas (1.3) and (1.4) are only suitable to the situation that $K_2$ is accelerated, unsuitable to the situation that $K_1$ is accelerated. The formulas (1.5) and (1.6) are only suitable to the situation that $K_1$ is accelerated, unsuitable to the situation that $K_2$ is accelerated.

In order to explain the relativity of length contraction, Einstein put forwarded the concept of relativity of simultaneity. The concept claims that if two events take place simultaneously at two different sites of an inertial frame, they would not take place simultaneously for another inertial frame with a relative speed. In this way, the relativity of length contraction is turned over to the relativity of simultaneity. However, for nonequivalent inertial reference frames, there is no the relativity of simultaneity. Though the readings of clocks located at the different points of an accelerated reference frames are different, the concept of simultaneity is still absolute. For nonequivalent inertial frames, if two events take place simultaneously for a reference frame, they also take place simultaneously for another reference frame, though the time's readings of the different clocks located at the different points of this reference frame may be different.

So according to this absolute and variable space-time theory, the real significance of the invariability principle of light's speed is that we can not make the speed of an object with rest mass reach and exceed light's speed in vacuum by the method of accelerating it. Only in this meaning, we can regard light's speed as a limit speed. Because the Lorentz transformation formulas still hold, the description forms of physics laws are still the same in the nonequivalent inertial reference frames.

## 5. The necessity and possibility of absolute resting reference frame

As shown in Fig. 1.4, let $K_0$ represent absolute resting reference frame. $K_1$, $K_2$ and $K_0$ are at rest in beginning and same rulers and clocks are defined for them. Then we accelerate $K_1$ and $K_2$ towards right sides simultaneously in accelerations $a_1$ and $a_2$ with $a_2 > a_1$ individually. When the speed of $K_1$ reaches $V_1$ and the speed of $K_2$ reaches $V_2$, stopping accelerating processes. After that, $K_1$ and $K_2$ move relative to $K_0$ in uniform speeds $V_1$ and $V_2$ with $V_2 > V_1$ individually. Suppose at a certain moment, the original points of three reference frame's coordinates just meet together. The readings of clocks at three frame's original points are adjusted to $t_0 = t_1 = t_2 = 0$ at this time. Meanwhile, a bind of light is send out along the $X$ axes. In light of the invariability principle of light' speed among non-equivalent inertial frames, the Lorentz transformations can be written as

$$x_1 = \frac{x_0 - V_1 t_0}{\sqrt{1 - V_1^2/c^2}} \qquad t_1 = \frac{t_0 - V_1 x_0/c^2}{\sqrt{1 - V_1^2/c^2}} \qquad (1.7)$$

$$x_2 = \frac{x_0 - V_2 t_0}{\sqrt{1 - V_2^2/c^2}} \qquad t_2 = \frac{t_0 - V_2 x_0/c^2}{\sqrt{1 - V_2^2/c^2}} \qquad (1.8)$$

The relative speed between $K_1$ and $K_2$ is

$$V = \frac{V_2 - V_1}{1 - V_1 V_2/c^2} \qquad (1.9)$$

The space-time transformation relations between $K_1$ and $K_2$ are

$$x_2 = \frac{x_1 - V t_1}{\sqrt{1 - V^2/c^2}} \qquad t_2 = \frac{t_1 - V x_1/c^2}{\sqrt{1 - V^2/c^2}} \qquad (1.10)$$

The invariability principle of light's speed between $K_1$ and $K_2$ still holds, but their relative speed should be defined by Eq.(9). Because of $a_2 > a_1$, $V_2 > V_1$ in this case, we should think that $K_2$ is accelerated relative to $K_1$, so that the rulers and clocks on $K_2$ contract absolutely comparing with that on $K_1$.

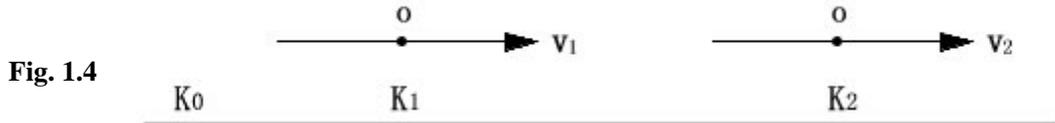

**Fig. 1.4**

On the other hand, as shown in Eq.(7), when accelerated $K_1$ moves in uniform speed $V_1$, the clocks located at different places dedicate different times. The observers in $K_1$ can regulate these clocks so that their time becomes the same. After that, the observers in $K_1$ would find that light's speed is different from $K_0$. There is no light's speed invariability again between $K_0$ and $K_1$ in this case. Then the observers in $K_1$ accelerate another reference frame $K_1'$ with same ruler and clock, and stop accelerating when $K_1'$ reaches a certain speed. After that, there exists the invariability of light's speed and the Lorentz transformation between $K_1$ and $K_1'$, but there exists no the invariability of light's speed and the Lorentz transformation between $K_1'$ and $K_0$ (as well as between $K_1'$ and $K_2$). In fact, all nonequivalent inertial reference frames on the earth are established in this way. We always take the earth as the resting reference fame, define the same rulers and clocks on it, then establish nonequivalent reference frames by the method of accelerating them and carry out physical experiments on them.

The discussion above would lead to a result, that is, the absolutely resting reference frame exists. As shown in Fig. 1.4, if we accelerate $K_1$ towards right side so that its speed becomes faster after it obtain a uniform speed $V_1$, the rulers in $K_1$ would become shorter, the clocks would become slower further and the mass of object would becomes greater. Oppositely, if we accelerate $K_1$ towards left side so that its speed becomes slower after it obtains a uniform speed $V_1$, the rulers in $K_1$ would become longer and the clocks would become quicker and the masses of objects would become smaller. When the speed of $K_1$ becomes zero relative to the absolutely resting reference frame $K_0$, the rulers are longest, the clocks are quickest and the masses of objects are smallest. But if we continue to accelerate $K_1$ towards left side, the rulers in $K_1$ would become short and the clocks would become slow and the masses of objects would become great again.

It is obvious that only the space-time of absolutely resting reference frame can be isotropy. The rulers are longest, the clocks are quickest and masses are smallest in the absolutely resting reference frame. No other frames have such characteristics. So from the angle of symmetry, such a reference frame exists, in which space-time is completely isotropy. We can find it by accelerating a reference frame in the different directions of space and observing the changes of ruler, clock and object's mass in the frame. If experimental results are isotropy at a certain state, the reference frame in this case can be regarded as the absolutely resting reference frame.

A simple method is provided here to determine the absolutely resting reference frame. Because object's moving mass is the smallest in the absolutely resting reference frame (equal to its rest mass), as long as we can find out a reference frame in which a object's mass is smallest, we can think that this reference frame is just the absolutely resting one. The concrete method is below. A chamber as shown in Fig. 1 is fixed in a rocket. An object is connected to a sensitive spring. Then we accelerate the rocket in a constant acceleration in different directions of space. The spring would be pulled long by the inertial force caused by the object. Measure the length's change of spring accurately by optical method. If we find that the pulled length of spring is the shortest at a certain moment, stopping accelerating. Then the rocket would move in a uniform speed in space. In this case, all inertial reference frames which are at rest with the rocket can be regarded as the absolutely resting reference frames.

The existence of absolutely resting reference frame coincides with the demand of modern cosmology and practical observations of astronomy. In the 1960's, astronomers found the special anisotropy of microwave background radiation[4]. If we take the reference frame, in which microwave background radiation is isotropic, as the absolutely resting one, astronomic observations showed that the earth reference frame was moving in a velocity $V \leq 300$ Meters/s towards to the directions of right ascension $1^h.5 \pm 0^h.4$ and declination $0.^02 \pm 7^0$. This velocity could be regarded as the velocity that the earth moves relative to the absolutely resting reference frame. In 1999, the anisotropy detector of microwave background radiation (WMAP) found anisotropy at higher precision. In 2002, physicists found the anisotropy of radio waves eradiated by radio galaxy in the earth's motion direction by using array radio telescopes (VLA)[5]. This kind of anisotropy is also considered to be caused by the Doppler effect of the earth's motion. Any more, the big-bang cosmology demands the existence of the absolutely resting reference frame. We can take the reference frame which is at rest with the original point of big-bang universe as the absolutely resting reference frame. The other reference frames produced by accelerating in the Big-bang processes would be considered to move relative it.

## 6. The equivalent inertial frames and possibility of surpassing light's speed in vacuum

We define the reference frames as the equivalent inertial reference frames among them the relative velocities exist primordially, not caused by accelerating processes. That is to say, there exist no relative resting states among equivalent inertial frames from beginning to end, so that we can not define same rulers and clocks for them at resting states. The definitions of rulers and clocks in equivalent inertial frames can be completely independent. These kinds of equivalent inertial reference frames, of cause, may be completely equivalent and have no any causality. So they are not restricted by the invariability principle of light's speed. In fact, all moving reference frames produced by the big-bang processes can not be equivalent inertial frames so that relative moving velocities among them can not exceed light's speed in vacuum. But if there are some celestial bodies which were not produced through the Big bang processes of our universe, relative velocities among our universal celestial bodies and these celestial bodies can be allowed to surpass light's velocity in vacuum.

On the other hand, for non-inertial reference frames, some apparent relative velocities may be allowed to surpass light's velocity when these velocities are not caused by accelerating processes. For example, the tangent velocities of all fixed stars exceed light's speed when we observe on the rotating earth. But we can not find that the fixed star's diameters become zero even imaginary number. If we build a uniformly rotating disc on the earth, as long as the angle speed of disc reaches $\omega \geq 1/480$, the tangent speed of the sum would reach and surpass over light's speed. But the diameter of the sum is almost unchanged for the observers on the disc. All these facts can not be explained by Einstein's special relativity. But it can be explained well by this paper's theory. Because the sum and fixed stars have not ever been accelerated in the process, their diameters do not contract. The fact is that the rotating disc has been accelerated, so the tangent length of disc contracts. So we only can say that the earth and disc are rotating, instead of the sun and fixed stars. By using the ruler in the rotating disc to measure, the diameters of the sum and fixed stars would become longer, instead of shorter, for the ruler placed on the disc becomes shorter. Though the change is very small when disc's tangent speed $V = r\omega << c$. Meanwhile, all observers on the earth and the sun or the fixed stars would have the same viewpoint about the motions and velocities. There is no relativity in this case. In fact, energy needed to accelerate a fixed star is completely different from that to accelerate a disc. This kind of asymmetry would cause some measurable effects.

It should be emphasized again that the invariability principle of light's speed only means that the speed of an object with rest mass can not reach and exceed light's speed in vacuum by means of the method of accelerating it. The reason to cause light's speed invariable is that the space-time scales of non-inertial reference frame change really in accelerating process, so that the observers in the non-inertial frame find that light's speed is still unchanged. In fact, so-called light's speed invariability is conditional. For example, in medium, light's speed is variable. It depends on medium's refractive index. When refractive index is big than 1, the speed of light in medium is smaller than its speed in vacuum. This is a well-known fact in physics. If some day we find that a certain medium's refractive index is small than 1, it needs not to surprise that light's speed in this medium would surpass its speed in vacuum. This result does not violate the invariability principle of light's speed among the nonequivalent inertial reference frames. At first, the speed of light is not produced by accelerating process, Secondly, the light moves in medium, instead of vacuum. In fact, some authors declaimed in 2000 that they had founded the phenomena of surpassing light's speed in vacuum in atom $S_c$ gas [6].

In sum, because velocity caused by accelerating process can not be relative, we have to abandon the principle of relativity, while the invariability principle of light's speeds in vacuum among nonequivalent inertial reference frames are still hold. We should establish space-time transformation theory based on the nonequivalent inertial reference frames. In this way, we can reach an absolute space-time theory with variable space-time scales. This kind of theory has no any contradiction in logic. It does not contradict with any present physical experiments and astronomic observations. Meanwhile, it can coincide with modern cosmology. So it can be considered as a really rational space-time theory to describe our physical world.

# Section 2　Rationality Problems of the Principles of Equivalence and General Relativity

### 1. Introduction
It is pointed out that the equivalent principle and the relativity concept of space-time are incompatible in logic. The current *Eötvös* type experiments only proves that inertial static mass and gravitational static mass are equivalent, without proving that inertial moving mass and gravitational moving mass are equivalent. By means of the dynamic effect of special relativity, it is proved that inertial forces and gravity are not equivalent locally. Meanwhile, by means of abundant facts and logical analysis, it is also shown that the principle of general relativity is actually untenable. There exists logical difficulty to explain the gravitational redshift of spectrum based on the equivalent principle of general relativity. It is pointed out that the hypotheses of space-time curve have no direct verification actually and is also unnecessary. The logic foundation of the Einstein's theory of gravitation has serious objection. We have to renovate our idea about space-tine and gravitation.

### 2. The equivalent principle is not consistent with special relativity
　　We point out at first that the equivalent principle is not consistent with special relativity. Einstein used the reference frame of rotational desk to show the principle of equivalence and concluded that the space-time of gravitational field was curved. As shown in Fig. 2.1, suppose that the desk $K_1$ with radium $r$ is at rest in beginning. Then let the desk turn around its center in a uniform angle speed $\omega$. According to length's contraction in special relativity, observers on the resting reference frame of ground $K_0$ would find that the perimeter of desk becomes $2\pi r\sqrt{1-\omega^2 r^2/c^2}$ owing to desk's rotation, but the radium of desk would not contract for there is no motion in the direction of radium. Therefore, the radio of perimeter and radium would be less than $\pi$, so that the observers on the ground would affirm that the space of rotating desk is non-Euclidean space.

　　The problem now is that what the observers resting on rotating desk think about. Einstein and almost textbooks of general relativity took a mistake here [7]. According to Einstein and the current textbooks, the rulers of observers on rotating desk would contract along the tangent direction of desc, but the perimeter of desk would not contract. So when observers on rotating desk used their rulers to measure the perimeter of desk, the perimeter becomes $2\pi r/\sqrt{1-\omega^2 r^2/c^2}$, longer than that observed on ground. But when they use their rulers to measure the radium of desk, the length is unchanged. So the ratio of perimeter and radium is big than $\pi$ for observers on rotating desk. This result is obviously wrong, for it is based on the premises that desk is at rest and observer and his ruler move. But in this case, desk rotates actually so that the perimeter of desk and the rulers of observers contract synchronously. So the observers who are at rest on desk can not find the change of desk's perimeter actually by using their rulers! That is to say, the observers in curved space could not find space's curvature actually if they only use their rulers to measure. It involves to the possibility problem of measurements in curved space-time as well as the rationality problem that we compare directly the calculating results in curved space-time with the experiments carried out in flat space-time. We will discuss them again in Second 4.

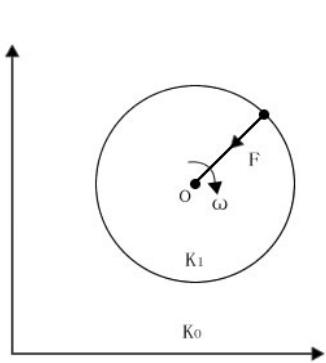 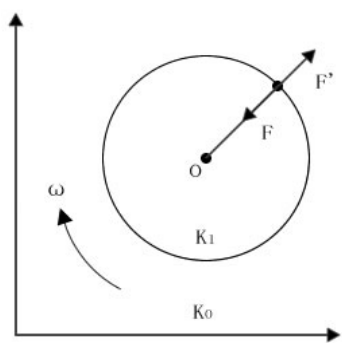 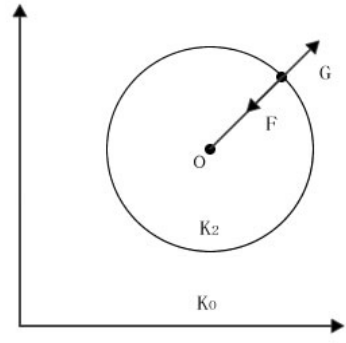

Fig. 2.1    Fig. 2.2    Fig. 2.3

Secondly, let's show that the principle of equivalence contradicts with the concept of space-time relativity. There are two problems here. The first is that according to the common understanding of special relativity, space-time's contraction is caused by relative speed, which has nothing to do with acceleration[8]. So according to space-time relativity, for the observers resting on rotating desk, $K_1$ is at rest but ground reference frame $K_0$ moves around $K_1$ as shown in Fig. 2.2. In this way, the observers resting on desk would think that the rulers on their hands are unchanged so that they can not conclude that space-time of desk is curved. Instead, they would think that the space-time of ground reference frame $K_0$ should be curved for $K_0$ is rotating around $K_1$. Of course, this result can not be accepted. Therefore, if the concept of space-time relativity holds, we can not conclude that the space-time of gravitational field is curved. Conversely, if the space-time of gravitational field is curved and the principle of equivalence holds, the principle of space-time relativity would be violated.

The second problem is that in light of general recognition of special relativity, space-time contract is a purely relative effect, having nothing to do with acceleration. But according to the equivalent principle, space-time contraction should be relative to force and interaction, not a purely relative effect.

It is obvious that the principle of equivalence contradicts with the principle of space-time relativity. Because the principle of equivalence is regarded as the foundation of the Einstein's theory of gravitation, it can be said that special relativity is not consistent with the Einstein's theory of gravity. Only way for us to get rid of this paradox is to give up the relativity of space-time and admit that space-time contraction is a kind of effect caused by accelerating processes. The rational result should be that the perimeter of rotating desk contracts and the radio of perimeter and radium is less than $\pi$. No matter the observers are on rotating desk or on ground, they viewpoints are the same. Only in this way, we can each a logically consistent theory of space-time and gravitation.

## 3. The non-equivalence of Inertial moving mass and gravitational moving mass

Let's discuss the problem of equivalent principle itself. The principle can be divided two parts, one is weak equivalent principle and another is strong equivalent principle. The weak equivalent principle indicates that gravitational mass is equivalent with inertial mass, or gravity is equivalent to inertial force locally. We first discuss the equivalence of gravitational mass and inertial mass. Then discuss the equivalence of gravity and inertial force, as well as the relation between curved space-time in gravitational field and non-inertial reference frame and so do. It will be shown by considering the dynamic effects of special relativity that the principle of equivalence is impossible actually.

At present, the experiments to prove the equivalence between inertial mass and gravitational mass is the so-called *Eötvös* type of experiments. It should be pointed out that this type of experiments only prove the equivalence between inertial static mass and gravitational static mass, but can not be used to prove the equivalence of inertial moving mass and gravitational moving mass actually. According to the second law of Newtonian, the force acting on an object is $F = m_{i0} a$. Here $m_{i0}$ is inertial static mass and $a$ is acceleration. In a uniform gravitational field, the formula of force can also be written as $F = m_{g0} g$. Here $m_{g0}$ is gravitational static mass and $g$ = constant is the intensity of gravitational field. So we have

$$a = \frac{m_{g0}}{m_{i0}} g \tag{2.1}$$

If gravitational static mass is equivalent with inertial static mass, the radio $m_{g0} / m_{i0}$ should be a constant for any material. In this way, the acceleration of an object falling in this uniform gravitational field would also be a constant. This result has an obvious defect, i.e., the object's speed would surpass light's speed in vacuum at last when the object falls in the gravitational field. In order avoid this problem, the dynamic relation of special relativity should be considered. We have

$$m_i = \frac{m_{i0}}{\sqrt{1 - V^2/c^2}} \qquad F = \frac{d}{dt} \frac{m_{i0} V}{\sqrt{1 - V^2/c^2}} = \frac{m_{i0} a}{(1 - V^2/c^2)^{3/2}} \tag{2.2}$$

Because the effect of velocity on gravitational moving mass $m_g$ is still unknown at present, we suppose

$$m_g = f(V) m_{g0} \tag{2.3}$$

The function $f(V)$ is waited to be determined. Suppose we still have $F = m_g g$, Eq.(2.1) becomes

$$a = g f(V) \left(1 - \frac{V^2}{c^2}\right)^{3/2} \frac{m_{g0}}{m_{i0}} \tag{2.4}$$

As long as $f(V) \neq (1 - V^2/c^2)^{-3/2}$, the acceleration would not be a constant again. When $V \to c$, we may have $a \to 0$. The falling speed would not surpass light's speed. In the *Eötvös* type of experiments, the moment acting on the hang bar is [9]

$$T = l_A a_T m_g^A \left( \frac{m_i^A}{m_g^A} - \frac{m_i^B}{m_g^B} \right) \tag{2.5}$$

Here $l_A$ is the length of bar, $a_T$ is the centrifugal acceleration on the level direction caused by the earth's rotation, $m_i^A$ and $m_i^B$ are the inertial masses, $m_g^A$ and $m_g^B$ are the gravitational masses of two different materials A and B individually. If the dynamic effect of special relativity is considered, by using formulas (2.2) and (2.3), Eq.(2.5) can be written as

$$T = \frac{l_A a_T m_{g0}^A}{\sqrt{1 - V^2/c^2}} \left( \frac{m_{i0}^A}{m_{g0}^A} - \frac{m_{i0}^B}{m_{g0}^B} \right) \tag{2.6}$$

It is obvious that $T$ has nothing to do with $f(V)$. As long as the radio $m_{i0}/m_{g0}$ does not change with different material, no matter what is the form of function $f(V)$, we always have $T = 0$. It means that the current experiments of the *Eötvös* types only verify the equivalence between gravitational static mass and inertial static mass, without verifying the equivalence between gravitational moving mass and inertial moving mass. The relation between them will be deduced in Section 4, showing that gravitational moving mass and inertial moving mass are nonequivalent.

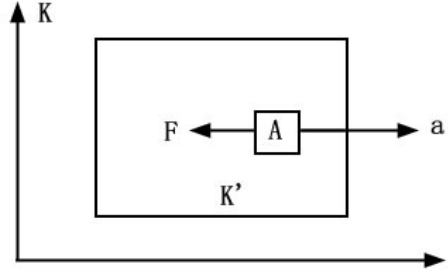

Fig. 2.4   Non-equivalence between gravitation and inertial force

On the other hand, the weak principle of equivalence can also be described as that gravitation is locally equivalent with inertial force. Besides rotating desk, Einstein used the ideal experiment of closed chamber to show the equivalence. In the experiments, observers in a closed chamber could not decide whether they were at rest in a uniform gravitational field, or were being accelerated with a uniform acceleration. This result is correct in the Newtonian theory, but can be proved to be incorrect after the dynamic effect of relativity is considered. As shown in Fig. 2.4, suppose that the chamber $K'$ is accelerated relative to resting reference frame $K$. There is an object $A$ with static mass $m_0$ is hanged on the ceiling of chamber through an elastic rope. When $K'$ is accelerated in a uniform acceleration, owing to the dynamic effect of special relativity as shown in Eq.(2.2), the inertial moving mass of object would become bigger and bigger. So the rope would be pulled longer and longer besides the Lorentz contraction. When the speed of chamber is great enough, the inertial force of object would surpass the bearing capacity so that the rope would be pulled apart. Because the breaking of rope is an absolute event, observers who are inside and outside chamber would observe it. The observers inside the chamber would affirm that they are accelerated, instead of resting at a static gravitational field. If the closed chamber was at rest in a static gravitational field, the rope would not be pulled longer and longer, and break at last. It is obvious that the equivalence between gravitation and inertial force is only tenable in the Newtonian mechanics. After the dynamic effect of special relativity is considered, gravitation and inertial force are not equivalent locally in general.

Besides, there are many other experiments to show the impossibility of weak equivalent principle. For example, when a closed chamber is at rest on a uniformly rotating desk, there exists the Coriolis force for a moving object, the inertial gyroscope would change its direction and charged objects would radiate photons. But in a static gravitational field, there are no such phenomena. Observers can distinguish both to be on rotating desk or at rest in a uniform gravitational field by theses phenomena. The results above show that it is improper to take the principle of weak equivalence as a fundamental principle of physics.

## 4. The equivalence of non-inertial frame and gravitational field on space-time bending

The equivalent problem of space-time bending between non-inertial frame and gravitational field is discussed below. At present, the time delay of gravitational field is considered as the result of equivalent principle, having nothing to do with the equation of gravitational field. Einstein used the ideal experiment of rotating desk to shown that the space-time of gravitational field was curved based on equivalent principle. We now discuss this problem. Because the clocks on rotating desk have velocities, clock's times on the desk would become slow. The relation between centripetal acceleration and the speed of circumference motion is $a = V^2/r$, we have time delay

$$dt = dt_0\sqrt{1 - \frac{V^2}{c^2}} = dt_0\sqrt{1 - \frac{ar}{c^2}} \qquad (2.7)$$

Here $dt$ is the time of clocks on rotating desk, $dt_0$ is the time of clocks on static ground. According to the principle of equivalence, as shown in Fig. 2.2., the reference frame of uniformly rotating desk is equivalent with a static gravitational field which does not change with time. In the figure, $F$ is the centripetal force acting on clock, $F'$ is inertial force. The equilibrium of two forces keeps the clock at rest on the desk. According to the equivalent principle, as shown in Fig. 2.3, $F$ should be regarded as a certain force just as a frictional force. This force balances gravitation $G$ so that the clock can be at rest in gravitational field without falling down. So as long as the equivalent principle is tenable, the time would become slow for a clock at rest in a gravitational field, determined by (2.7). Because rotating desk can be regarded as a system with centripetal force, the force acted on the object at rest in the force field can be written as $G = F' = F = ma = -km/r^2$. Substitute it into Eq.(2.7) we get time delay

$$dt = dt_0\sqrt{1 - \frac{ar}{c^2}} = dt_0\sqrt{1 - \frac{k}{c^2 r}} \qquad (2.8)$$

But if a non-inertial system moves in the field of centripetal force, the force acted on the object fixed in the frame would be $F = ma = -km/r^2$. Suppose the speed of the frame is zero when $r \to \infty$, according to classical theory of the Newtonian mechanics, we have relation

$$\frac{1}{2}mV^2 - \frac{km}{r} = 0 \qquad \text{or} \qquad V^2 = \frac{2k}{r} \qquad (2.9)$$

We can fix a clock on the particle. By substitute Eq.(2.9) into the first equation of Eq.(2.7), we can obtain the time delay of clock fixed on the particle

$$dt = dt_0\sqrt{1 - \frac{2k}{c^2 r}} \qquad (2.10)$$

The result is different from Eq.(2.8). It means that the time delays of the circle motion and the accelerating motion of straight line are different. Time delay depends on the forms of motions. But in general relativity, such distinction does not exist. The time delay is always represented by Eq.(2.10) in which $k = GM$.

In the 60's last century, time delay based on Eq.(2.10) was verified by the experiments of atomic cesium clocks circling the earth. The red shift of spectrum also verified the correction of Eq. Eq.(2.10). Now these experiments are considered as the proof that the existence of gravitational field would cause time delay. However, in the experiment of atomic cesium clocks circling the earth, clocks moved around the earth, in spite of moving along the radium direction of the earth. According to the equivalent principle, time delay should obey Eq.(2.8), in spite of Eq.(2.10).

Because time process is determined by the frequency of atomic libration, the fact that the frequency of atomic libration becomes slow means that time process becomes slow. It can be said that the hypothesis that gravitation would cause time slow has been verified by the experiments of spectrum red shift of and atomic cesium clocks circling the earth. Because time delay shown in Eq.(2.10) is not a quantity which can be measured directly, only by the indirect experiments just as the red shift of spectrum, we can decided it. At present, all experiments about it are carried out in weak fields of gravitation, it is still be worth of suspicion that if the formula would be correct in strong fields. In the forth section of this paper, we will provide a more rational formula about the effect of gravitation on the atomic libration or light's frequency based on the description of gravity in flat space-time. The big red shift of quasars can be explained well by the formula.

Then let's discuss the problem of space bending based on the equivalent principle. According to the current theory, when the angle speed of rotating desk is $\omega$, the space metric of desk can be written as [7]

$$d\sigma^2 = dr^2 + \frac{r^2}{1 - r^2\omega^2/c^2} d\varphi^2 + dz^2 \qquad (2.11)$$

It means that length contraction takes place along the tangent direction of desk. Because there is no speed along the radium direction, there exists no length contraction along this direction. On the other hand, according to the sphere symmetrical solution of the Einstein's equation of gravity, the space part of Schwarzschild metric can be written as

$$d\sigma^2 = \left(1 - \frac{2GM}{c^2 r}\right)^{-1} dr^2 + r^2 \sin^2\theta \, d\varphi^2 + r^2 d\theta^2 \qquad (2.12)$$

Because the slice of sphere is just the desk, so the space bending in both situations should be the same. In fact, we can let $dz = 0$ in Eq.(2.11) for the thin desk and can let $\theta = \pi/2$ and $d\theta = 0$ in Eq. (2.12) for the slice of sphere. In this case, Eqs.(2.11) and (2.12) should describe the same space metric. However, we can see that Eqs.(2.11) and (2.12) are not consistent in this case. In the Schwarzschild metric, length contraction takes place in the direction of radium, which is just the acting direction of gravitation. But in the metric of rotating desk shown in Eq.(2.11), length contraction takes place in the tangent direction of desk, instead of the direction of inertial force. It indicates that space bending caused by non-inertial force may be different from that caused by gravitation, if gravitational fields would also cause space bending really. Unfortunately, this obvious difference has not caused person's attention up to now. When a non-inertial frame moves around a center, the centripetal acceleration can be written as $a = r\omega^2 = k/r^2$. Substitute it into Eq.(2.11), we have

$$d\sigma^2 = dr^2 + \frac{r^2}{1 - k/rc^2} d\varphi^2 + dz^2 \qquad (2.13)$$

If the non-inertial frame moves along the direction of centripetal force, we have relation $V^2 = 2k/r$. According to the Lorentz contraction, the metric of non-inertial frame can be written as

$$d\sigma^2 = \left(1 - \frac{2k}{c^2 r}\right)^{-1} dr^2 + r^2 \sin^2\theta \, d\varphi^2 + r^2 d\theta^2 \qquad (2.14)$$

Eq.(2.14) coincides with Eq. (2.12), but not with Eq.(2.13). So according to the equivalent principle, the space scalar of a reference frame at rest in gravitational field should be equivalent with the non-inertial reference frame which moves along the radium direction of centripetal force, in spite of that moves around the centripetal point.

In fact, we lack of direct proofs to prove that gravity would cause space bending. What we have now is only indirect proofs. We only prove it by calculating object's motions in the gravitational field of the sun and can obtain more accurate results. The result is equal to object's motions in curved space-time. But all of them are indirect proofs. This is different from time delay in gravitational field, though as shown in the later discussion in this paper, time delay in gravitational field can be deduced by a different method, in spite of the equivalent principle. So it is unproved that gravitation would cause space bending. In fact, we can establish gravitational theory in flat reference frame without introducing the hypothesis that gravitation would cause space bending. Though the Einstein's theory of gravity has obtained great successes, too many space-time singularities appear in the theory. Within the singular region, all physical laws lose efficacy so that we can not research their natures. The essential reason is owing to that the theory is established in curved space-time. If gravitational theory can be established in flat space-time, owing to the fact that the space-time curvature is always zero, we have no any singularity problem again. As shown below, after the theory of gravitation is established in flat space-time, all space-time singularities are transformed into the infinite points of gravitations. This kind of infinites would appear in any theory based on the mode of point particles which is not worth to be surprise. In the Newtonian's classical gravitational theory, when $r \to 0$ gravitation would becomes infinite. But we do not worry about this kind of infinite, for its nature is completely different from space-time singularities appearing in general relativity. This problem will be discussed in detail in Second 4.

### 5. The existing problem to explain gravitational redshift by the equivalent principle

It is proved below that there exists logical difficulty to explain the gravitational redshift of spectrum based on general relativity. In general relativity, gravitational redshift is explained by the equivalent principle, which has nothing to do with the equation of gravitational field. Suppose that when there exist no gravitational field, the proper frequency of light emitted by an atom is $v_1$ and the proper wave length is $\lambda_1$. Let's fist discuss the result when this atom is moved into the point $r_1$ in the gravitational field. Suppose that when the light emitted by the atom in gravitational field reaches the observers who are at rest outsider the gravitational field with $r_2 \to \infty$, the observed frequency and wave becomes $v_2$ an $\lambda_2$. Meanwhile, we suppose that light's speed is unchanged, no matter inside or outside the gravitational field, we have $c = \lambda_1 v_1 = \lambda_2 v_2$. According to the equivalent principle of general relativity, in the gravitational field with spherical symmetry, time contraction formulas are

$$d\tau_1 = dt\sqrt{1-\frac{\alpha}{r_1}} \qquad d\tau_2 = dt\sqrt{1-\frac{\alpha}{r_2}} = dt \qquad (2.15)$$

In which $d\tau_1$ is the proper time and $dt$ is the coordinate time. It has been proved that for light's motion in the static gravitational field, $dt$ is unchanged [15], so the formulas of time contraction indicates that the clocks at rest in gravitational field would become slow. In general relativity, or in most of text books and documents, the relation between the libration frequency of atom and proper time is supposed as [3] [15]

$$\frac{v_1}{v_2} = \frac{d\tau_2}{d\tau_1} \qquad (2.16)$$

The redshift is defined as

$$Z = 1 - \frac{\lambda_1}{\lambda_2} = 1 - \frac{v_2}{v_1} \qquad (2.17)$$

By considered the formula above, we can get

$$Z = 1 - \frac{\nu_2}{\nu_1} = 1 - \frac{d\tau_1}{d\tau_2} = 1 - \sqrt{1 - \frac{\alpha}{r_1}} > 0 \qquad (2.18)$$

In the weak field with $\alpha/r_1 \ll 1$, we have $Z = \alpha/(2r_1)$ which coincides with experiments.

Though there is no problem in mathematics, there exists a logic difficulty to explain gravitational redshift here. Because of $d\tau_1 < d\tau_2$ according to (2.15), we have $\nu_1 > \nu_2$. It indicates that when a free atom outside the gravitational field is moved into the gravitational field, its libration frequency would become great, or the frequency of light emitted by the atoms in gravitational field would becomes great. If light's speed is unchanged in gravitational field, it means that the wave length of light emitted by the atoms in gravitational field would become small with $\lambda_1 < \lambda_2$. Therefore, there exist three possibilities when the light arrives at the observers who are at rest outside the gravitational field.

The fist possibility is that observers outside the gravitational field observe purple shift, in stead of redshift. According to general relativity, photon would not is acted by force when it moves along the geodesic in gravitational field without the concept of potential energy, so photon's energy can be written as $E_1 = h\nu_1$ when it is at the point $r_1$ of the field. Suppose photon's energy is unchanged when it moves in the field, its frequency is also unchanged. Because light's speed is unchanged, light's wave is also unchanged in the propagation process of light in gravitational field. So when the light arrives at the observers at rest outsider the field with $r_2 \to \infty$, its frequency and wave length are also unchanged. We still have $\lambda_1 < \lambda_2$ and $\nu_1 > \nu_2$. The observers outside the gravitational field would observe light purple shift, in stead of redshift. Of cause, this result is inconsistent with practices.

The second possibility is that observers outside the gravitational field can not find the shift of wave length. Because light is also a wave, its libration frequency would also change at the different point of gravitational field if time contraction is different at different point. When the strength of field decreases, light's frequency would decrease and its wave length would increase. When the light emitted by the atom in gravitational field reaches the observers outsider the gravitational field with $d\tau \to d\tau_2 = dt$, its wave and frequency would become original $\lambda_1$ and $\nu_1$ again. That is to say, the observer outsider the gravitational field would not find light's redshift. The precondition is that light's frequency would change in gravitational field. Because there is no the concept of gravitational potential in general relativity, photon has no potential energy. The result indicates that the energy conservation law is violated when photon moving in gravitational field.

At last, we discuss the third possibility. When the atom outsider the gravitational field with proper wave length $\lambda_1$ and frequency $\nu_1$ is moved into the point $r_1$ of gravitational field, according to the observer who is at rest in the field near the atom, the wave length and frequency of the atom are still $\lambda_1$ and $\nu_1$. Or we suppose that atom's wave length and frequency are originally $\lambda_1$ and $\nu_1$ without considering how the atom is placed in the gravitational field. Thus, when the light with wave length $\lambda_1$ and frequency $\nu_1$ emitted by the atom travels through gravitational field and arrives the observers outside the field $r_2 \to \infty$, its wave length and frequency become $\nu_2$ and $\lambda_2$. We have $\nu_1 > \nu_2$ and $\lambda_1 < \lambda_2$. In the current general relativity, the calculation of spectrum redshift is just based on this hypothesis. But in this case, there exist two problems here. At first, we have supposed that the libration frequency of atom is its proper frequency $\nu_1$ when the atom is located at the point $r_1$ in gravitational field. In order to obtain the result $Z > 0$, we suppose $d\tau_1 = dt\sqrt{1 - \alpha/r_1}$ again. Because of $\nu_1 \sim 1/d\tau_1$, it contradicts with the precondition that the libration frequency of atom is its proper frequency $\nu_1$. The atom's proper frequency is that measured by observers who are at rest together with the free atom without gravitational interaction. The proper frequency does not change in gravitational field. It does not obey the relation $v_1 \sim 1/d\tau_1$. Secondly, the hypotheses that atom's frequency does not change in gravitational field should be questioned, for it is inconsistent with the principle of equivalence. So the third possibility is also impossible actually.

It can be said that the source of logic difficulty is relative to the definition of (2.16). The formula is actually baseless. Only for the purpose to obtain the result of (2.18), we introduced it. Because time becoming slow indicates the libration frequency becoming mass, the real relation between libration frequency and proper time would be

$$v_1 \sim d\tau_1 = dt\sqrt{1-\frac{\alpha}{r_1}} \qquad v_2 \sim d\tau_2 = dt\sqrt{1-\frac{\alpha}{r_2}} = dt \qquad (2.19)$$

In fact, in some textbook and document of general relativity, the formula (2.19) is used [4] [7]. In light of this relation, we have no logic difficulty to explain gravitational redshift. We can suppose that when an atom with proper frequency $v_2$ and proper wave length $\lambda_2$ is moved into the point $r_1$ of gravitational field, its proper frequency and proper wave length become $v_1$ and $\lambda_1$. Because of $d\tau_1 < d\tau_2$, we have $v_1 < v_2$ and $\lambda_1 > \lambda_2$, i.e., the libration frequency of atom in gravitational field becomes small and its wave length becomes longer. Because photon is not acted by force in gravitational field according to general relativity, photon has no potential energy. Owing to energy conservation law, photon's energy $E = hv_1$ is unchanged when it moves in gravitational field. So when the light emitted by the atom located in gravitational field arrives at the observers outsider the field, its frequency and wave length are still $v_1$ and $\lambda_1$, i.e., redshift is observed.

It can be said that the difficulty of explanation above is relative to the hypotheses (2.16). However, it is untrue. By considering the fact that time becoming slow indicates the libration frequency of atom becoming small actually, the real relation between frequency and proper time should be

$$v \sim d\tau = dt\sqrt{1-\frac{\alpha}{r}} \qquad (2.19)$$

The relation has actually been used in same textbooks [4] [7]. Suppose when the atom with proper frequency and proper wave length $v_1$ and $\lambda_1$ is placed into the point $r_1$ of gravitational field with spherical symmetry, its frequency and wave length becomes $v_2$ and $\lambda_2$. According to (2.19), we have

$$v_2 \sim d\tau_1 = dt\sqrt{1-\frac{\alpha}{r_1}} \qquad v_1 \sim d\tau_2 = dt\sqrt{1-\frac{\alpha}{r_2}} = dt \qquad (2.20)$$

It indicates $v_1 > v_2$ and $\lambda_1 < \lambda_2$, i.e., the frequency becomes small and the wave length becomes longer for the atom in the gravitational field. According to general relativity, photon is not acted by force without the concept of potential. Due to the law of energy conservation, photon's energy $E = hv_2$ is unchanged in gravitational field, so after arriving at the observers outsider the gravitational field, light's frequency and wave length are still $v_2$ and $\lambda_2$. By the redshift definition of (2.17), we have

$$Z = 1-\frac{\lambda_1}{\lambda_2} = 1-\frac{v_2}{v_1} = 1-\sqrt{1-\frac{\alpha}{r_1}} > 0 \qquad (2.21)$$

This kind of explanation seems rational, but the same problem mentioned above still exists. Because light is also a wave, its libration frequency would also change at the different point of gravitational field if time contraction is different at different point. When the atom is placed into the gravitational field, we declaim that its frequency and wave length would change. When light emitted by the atom moves in the same field, how can we declaim that its frequency and wave length would not change? This is illogical. When the strength of field decreases, light's frequency would increase and its wave length would decrease. So when the light with frequency $\nu_2$ and wave length $\lambda_2$ arrives at the observers outsider gravitational field, its frequency and wave length would become $\nu_1$ and $\lambda_1$ again. The result is that the observers can not find light's redshif. Meanwhile, because light's frequency changes, according to general relativity, the law of energy conservation would be violated. This is also unacceptable.

So situations are quite confused for the discussion of gravitational rdshift in the current general relativity. It can be said that we can not explain the gravitational redshift of spectrum in a rational logic based on the equivalent principle of general relativity. It can be pointed out that the main reason of logic difficulty to explain gravitational redshift is that we suppose light's speed is unchanged in gravitational filed. As proved in Section 4, in light of the Schwarzschild solution of the Einstein's equation of gravitational field strictly, light's speed would change when it moves in gravitational field with spherical symmetry. For the observers who are at rest in the field, light's speed is $V_c = c(1 - \alpha/r) < c$. Meanwhile, photon would be acted by repulsion force, in stead of gravitation. So photon's potential energy is positive. By considering these two factors, we can rationally explain the phenomena of gravitational redshift. The detail will be discussed in Second 4.

As we known that there are only two effects to cause the redshift of spectrum at present, one is the Doppler effect and another is the gravitational effect. The Doppler redshift can be deduced from special relativity and has been researched very well on theory and experiments. Relatively, the research of gravitational redshift is not abundant. So we should take the Doppler shift as background to study gravitational redshift. Similarly, suppose that $\nu_1$ and $\lambda_1$ are the proper frequency and wave length of light, or the frequency and wave length observed by the observers who are at rest together with the source of light, $\nu_2$ and $\lambda_2$ are the frequency and wave length measured by the observers who are located at the resting reference frame. The original definition of the Doppler shift in special relativity is

$$Z = \frac{\lambda_2}{\lambda_1} - 1 = \frac{\nu_1}{\nu_2} - 1 \quad (2.22)$$

It is obvious that the definition is different from that of general relativity shown in (2.17). If the definition (2.18) is used, by the same relations $d\tau_1 = dt\sqrt{1-\alpha/r_1}$ and $d\tau_2 = dt\sqrt{1-\alpha/r_2} \to dt$ as well as $\nu_1 d\tau_1 = \nu_2 d\tau_2$, what we get is

$$Z = \frac{\lambda_2}{\lambda_1} - 1 = \frac{\nu_1}{\nu_2} - 1 = \frac{d\tau_2}{d\tau_1} - 1 = \frac{1}{\sqrt{1-\alpha/r_1}} - 1 > 0 \quad (2.23)$$

Under the condition of weak field with $\alpha/r_1 << 1$, we also have $Z = \alpha/(2r_1)$. But in general situations, the results are different. Under the extreme condition with $\alpha/r_1 \to 1$, we have $Z = 1$ according to (2.18), but we have $Z \to \infty$ according to (2.23). The results are completely different. Which one is alright? This problem has not been seriously treated in general relativity. In order to be consistent with the Doppler shift, the formula (2.23) is also taken as the original definition of gravitational redshift in the paper. Meanwhile, we think that the relation between the libration frequency and proper time shown in the formula (2.19 is alright. Only based on these facts, we can explain gravitational redshift logically and rationally.

# 6. The impossibilities of strong equivalence principles and general relativity

The strong equivalent principle affirms as that by taking the local inertial reference frame at any space-time point in a gravitational field, the forms of nature laws are the same as they are in the reference frames in which the Descartes reference frame is taken without gravitational field and acceleration. Or speaking directly, in a local reference frame which is falling freely in a gravitational field, gravitation would be eliminated. This seems to be a well-known experimental fact. Then, is the strong equivalent principle tenable? The following analysis shows that this is only an apparent feeling of mankind and untenable actually.

In order to explain this result, we need to establish the concepts of "uniform force" and "non-uniform force". So-called "uniform force" indicates that the force acted on an object is distributed uniformly over all parts of the object. "Non- uniform force" indicates that the force is only acted on a part of the object. It is obvious that gravitation is "uniform force", but the force acted on an observer who is at rest in non-inertial reference frame is non- uniform force. For example, when an observer stands on an accelerated train, the force acted on him actually acts on his feet and the feet draw the body of observer moving forward. The other parts of body, not being acted directly by the force, are still in internal state, so that the body moves backward. In this way, the so-called inertial force is caused. In an accelerating elevator, the force is also only acted on observer's feet directly. Then the force is contributed to whole body through feet. For observers on rotating desk, friction force between feet and desk is need so that the observer can stand on desk. The friction force is also acted on feet directly. All of these forces are non-integral forces. It is just owing to the action of the non-uniform force, non-equilibrium or internal stress is caused in observer's body so that observer can feel the existence of accelerating motion. It should be pointed out that "non-uniform force" is only a macro-concept. The essence of "non-uniform force", such as friction force, is electromagnetic force. From micro-angle, however, all interactions between micro-particles are uniform force.

It is clear that the force acted on an object in uniform gravitational field is uniform force. The inertial force caused by non-inertial motion is also uniform force in general. The problem is now that when a non-uniform force is distributed over whole body of observer who is at rest in a non-inertial reference frame, what would happen? As shown in Fig. 2.1, the centripetal force acted on the feet of observer at rest on rotating desk is friction force actually. If this friction force is distributed over each part of observer's body, the observer would not fell this force's action. We can see in Fig.2.2 from the angle of non-inertial reference frame, this uniform friction force would be counteracted by the uniform inertial force. In this case, the observer who is at rest on rotating desk seems to be at rest in a static inertial reference frame without any force's action, though he is actually accelerated. He may also think that he himself is falling in a gravitational field, just as an astronaut travels in the orbit of cincturing earth.

So it is only a subjective feeling that observer was not acted by force when he fell freely in a uniform gravitational field. The true is that the force acted on observer's body is uniform force which is distributed over observer's body uniformly. In this way, no internal stress is produced in observer's body so that he can not fell the action of gravitational force and the existence of acceleration. Physiology tells us that felling is caused by non-uniformity. No felling does not indicate no changing. If changing is uniform and slow, there would be no felling. If proper methods are used, surpassing subjective felling, the observer would find himself to be being accelerated when he falls freely in a uniform gravitational field. For example, a charged object would radiate photons when it falls freely in a gravitational field, but the object would not when it is at rest in inertial reference frame. So only by taking a charged object with himself, the observer would judge whether he is falling freely in a uniform gravitational field or at rest in inertial reference frame. Only by this fact, the principle of strong equivalence has been proved untenable. More important is that by comparing with static inertial reference frame, moving ruler's length would contract, moving clock's time would become slow and moving object's mass would increase absolutely. The forms of nature laws are different for both the freely falling reference frames in gravitational fields and the Descartes inertial reference frame.

Einstein denied the existence of absolutely resting reference frame by means of establishing special relativity. In special relativity, all inertial reference frames were considered to be equivalent for the description of nature phenomena. After that, he introduced the principle of general relativity again to eliminate the superiority of inertial reference frame. In general relativity, all reference frames with arbitrary moving forms were considered to be equivalent for the description of nature phenomena. In mathematics, the principle of general relativity can be described as that the basic forms of motion equations should be covariant, or the line element of arbitrary reference frames must satisfy relation $ds^2 = $ constant. This demand is rational, for it actually indicates that light' speed in vacuum is still invariable under the condition of non-inertial coordinate transformations. But it does not mean that the forms of forces acted on objects are the same. As we known, in the common three dimension space, the motion equations can be transformed into other forms by introducing arbitrary coordinate transformations without bring any physical effect, for there are no any force or acceleration is introduced in this case. But in the four dimension space-time, the situation is completely different. In the four dimension space-time, if the transformation is linear just as the Lorenz transformation, a uniform velocity, instead of force or acceleration, would be introduced. But if the transformations are arbitrary, accelerations or non-inertial forces would be introduced in motion equations. In this way, there still exists a superior reference frame actually, i.e., inertial reference frame. In inertial reference frames, the forms of motion equations are simplest without the existence of inertial force.

In the Einstein's theory of gravity, non-inertial motions are considered to be equal to gravitational fields. At present, based on the principle of general relativity, it is thought that when the coordinate transformation is carried out for a solution of gravitational field's equation, the solution with new form would still be effective to describe the original gravitational field. However, this is impossible. If non-inertial reference frame is equal to gravitational field, a special non-inertial reference frame can be only equal to a special gravitational field. So when a non-inertial reference frame is transformed into another, it means that a gravitational field is transformed into another one with different nature. Because inertial forces are considered to be equal to gravities, the general transformation in the space-time of four dimensions would change gravitational fields. For a certain gravitational field, if the coordinate transformation is carried out, the form of field's equation and the solution would change. The result means that the original field has been transformed into new gravitational field, and new solution would not be original one. So a certain gravitational field can only correspond to a certain metric. Arbitrary coordinate transformation in the four dimension's space-time is not allowed, unless in new reference frame the same result can be obtained. However, this is impossible in general. For example, we can not re-calculate the perihelion precession of Mercury as well as other problems in the Lemaite or Kruskal coordinate systems and get the same result as we do in the Schwarzschild coordinate system. In fact, the reason why the energy-momentum tensors of gravitational field can not be defined well at present is actually owing to the existence of strong equivalent principle. According to the principle, we can introduce same local inertial reference frames through coordinate transformations, in which gravitational fields would be eliminated and the energy-momentum tensors of gravitational fields become zero. However, according to the law of energy-momentum conservation, a static system's energy and momentum should be constants. How can they be canceled only by coordinate transformations?

Therefore, it can be said that no relativity and arbitrarily are allowed in the description of gravity. The description of gravitational theory demands absolution. The principle of general relativity is untenable. The demand that the forms of motion equations should be covariant in arbitrary reference frames is only a basic restriction for the transformations of motion equations. It does not indicate that the descriptions of physical processes are the same in arbitrary reference frames. In fact, only in inertial reference frames, the basic forms of motion equations can be the same. And as mentioned in Section 1, only in the absolutely resting reference frame, the descriptions of physical processes are the simplest and most symmetrical.

At last, we should point out that a free falling of reference frame in gravitational field only means that the reference frame is accelerated. The force acted on the reference frame is only special one---- gravity. The result is that relative to resting reference frame, in the accelerated reference frame, length would contract and time would delay and mass would increase. All of those are the normal effects of special relativity. Therefore, as long as we know the interaction form between gravitational field and material, we can describe object's motion in gravitational field by the dynamic method of special relativity. The description is in flat space-time as shown in Section 4.

It is obvious that though the Einstein's theory of gravity has reached great success, there are many problems existing in its logical foundation and theoretical configuration. We should renovate our ideas about the essence of space-time and gravity.

# Section 3    Singularities Appears in the Gravitational Fields of Thin Circle and Double Spheres as well as the Rationality Problem of the Einstein's Equations of Gravitational Fields

**1. Introduction**

Though the Einstein's theory of gravity obtained great success, we have only four experiments to verify for the simplest solution of gravitational fields with mass static spherical symmetry distribution up to now actually. Speaking strictly, we only prove it in the weak gravitational field of the sum. As for other solutions of the Einstein's equations of gravitational fields, we can neither obtain experimental supports nor find corresponding physical systems for them. Facing so many forms of material distributions and comparing with so many experimental verifications of the Newtonian theory of gravity and quantum mechanics as well as special relativity, it is far not enough for the Einstein's theory of gravity.

By means of the coordinate transformations of the Kerr solution with double parameters and the Kerr-Newman solution with three parameters of the Einstein's equation of gravitation field with axial symmetry, the gravitational field equation's solutions for the static mass distributions of thin loop and double spheres can be obtained. The results show that no matter what are the masses and densities of loop and double spheres, space's curvatures in the centre points of loop and the double sphere's connecting line are infinite. The singularity points are completely exposed in vacuum and space curvatures nearby the singularity points and the surface of loop and double spheres are also highly curved even though the gravitational fields are very weak. It is obvious that all of them are actually impossible. The results show that the space-time singularities appearing in the Einstein's theory of gravity are not owing to high density and huge mass's distributions. They are actually caused by the Einstein's theory itself and have nothing to do with real world. The so-called black holes, white holes and wormholes with space-time singularity are something illusive, not existing in nature actually. All of these results indicate that the Einstein's theory of gravity can not bet a universally suitable one. Physicists would be clear-headed for the Einstein's theory of gravity. It is unadvisable for physicists to lose their judgment ability only by the great authority of Einstein and the beautiful form of the theory.

## 2. Thin circle's static gravitational field with axial symmetry

The gravitational problem of mass thin loop distribution is discussed at first. As shown in Fig. 3.1, a thin loop with mass $M$ and radium $b$ is placed on the x-y plane. The centre of loop is located in the original point of coordinate system. The loop is thin enough so that its cross section can be regarded zero comparing with its perimeter. It will be seen later that if the cross section of loop is not zero, the result is also the same in essence. Because the static mass distribution of thin loop is with axial symmetry, in light of the Einstein's theory of gravity, the metric tensor of gravitational field has nothing to do with time $t$ and coordinate $\varphi$, so the four dimension line element can be written as

$$ds^2 = g_{00}(r,\theta)dt^2 + g_{11}(r,\theta)dr^2 + g_{22}(r,\theta)r^2 d\theta^2 + + g_{33}(r,\theta)r^2 \sin^2\theta d\varphi^2 \qquad (3.1)$$

If space-time is flat, we have $g_{\mu\nu}=1$. By taking the coordinate transformations $t=t'$, $\varphi=\varphi'$, $r \to r'=r'(r,\theta)$, $\theta \to \theta'=\theta'(r,\theta)$, we can rewrite Eq.(3.1) as

$$ds^2 = g'_{00}(r',\theta')dt'^2 + g'_{11}(r',\theta')dr'^2 + g'_{22}(r',\theta')\ r'^2 d\theta'^2$$

$$+ g'_{33}(r',\theta')\ r'^2 \sin^2\theta' d\varphi'^2 + g_{12}(r',\theta')r'dr'd\theta' \qquad (3.2)$$

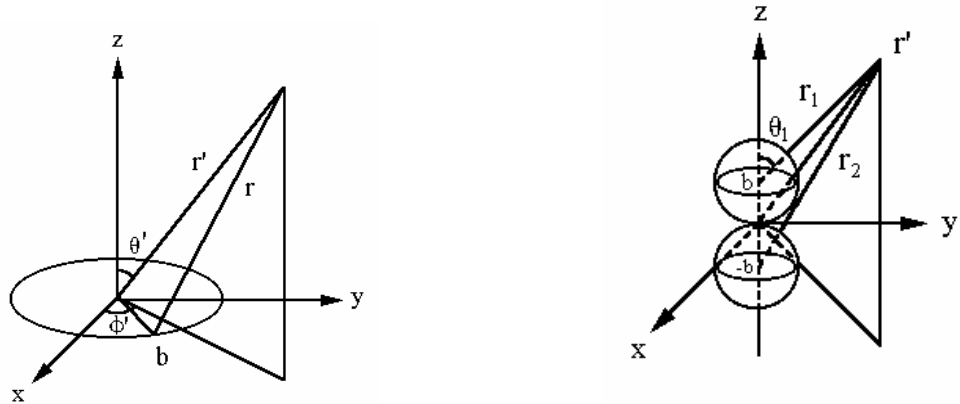

Fig. 3.1 Axial symmetry distribution of mass loop    Fig. 3.2 Axial symmetry distribution of double spheres

Two formulas above can be used to describe the gravitational field of thin loop. By putting the metrics above into the Einstein's equation of gravitational field, we can obtain the concrete forms of $g_{\mu\nu}$ in principle. But it is difficult to solve the equation of gravitational fields directly based on Eq.(3.1) or (3.2). On the other hand, there are two independent parameters $M$ and $b$ for the axial symmetry distribution of thin loop. There is also a ready-made solution of gravitational field's equation with axial symmetry and double parameters, i.e., the Kerr solution. If the solutions of the Einstein's equations of gravitational fields are unique, we can obtain the solution of static mass distribution of thin loop by means of the coordinate transformations of the Kerr solution. We have no other selection besides it. This method is discussed below. The Kerr solution is[10]

$$ds^2 = \left(1 - \frac{2\alpha r}{r^2 + \beta^2 \cos^2\theta}\right)dt^2 - \frac{r^2 + \beta^2 \cos^2\theta}{r^2 + \beta^2 - 2\alpha r}dr^2 - \left(1 + \frac{\beta^2}{r^2}\cos^2\theta\right)r^2 d\theta^2$$

$$-\left[1 + \frac{\beta^2}{r^2} + \frac{2\alpha\beta^2 \sin^2\theta}{r(r^2 + \beta^2 \cos^2\theta)}\right]r^2 \sin^2\theta\, d\varphi^2 + 2\frac{2\alpha\beta \sin\theta}{r^2 + \beta^2 \cos^2\theta} r\sin\theta\, dt d\varphi \qquad (3.3)$$

At present, the Kerr metric is regarded as the solution of a revolving sphere's gravitational field in which parameters $\alpha = GM$, $\beta = J/M$ is unit angle momentum ($c = \hbar = 1$). As for the static mass distribution of thin loop, the meanings of parameters $\alpha$ and $\beta$ are discussed below. Because the metric (3.3) contains a cross item $dt d\varphi$ relative to time, it is a solution of dynamic state. In the static mass distribution problem, this item does not exist. We can remove it by the diagonalization of metric tensors. Because only the items relative to $dt$ and $d\varphi$ should be taken into account, we can let

$$\begin{bmatrix} g_{00} & g_{30} \\ g_{03} & g_{33} \end{bmatrix} = \begin{bmatrix} 1 - \frac{2\alpha r}{r^2 + \beta^2 \cos^2\theta} & \frac{2\alpha\beta \sin\theta}{r^2 + \beta^2 \cos^2\theta} \\ \frac{2\alpha\beta \sin\theta}{r^2 + \beta^2 \cos^2\theta} & -1 - \frac{\beta^2}{r^2} - \frac{2\alpha\beta^2 \sin^2\theta}{r(r^2 + \beta^2 \cos^2\theta)} \end{bmatrix} \qquad (3.4)$$

Form the eigen equation

$$\begin{bmatrix} g_{00} - \lambda & g_{03} \\ g_{30} & g_{33} - \lambda \end{bmatrix} = 0 \qquad (3.5)$$

we can get

$$\lambda_1 = \frac{1}{2}(g_{00} + g_{33} + \sqrt{(g_{00} - g_{33})^2 + 4 g_{03} g_{30}})$$

$$= \frac{1}{2}\left[\left(2 - \frac{2\alpha r}{r^2 + \beta^2 \cos^2\theta} + \frac{\beta^2}{r^2} + \frac{2\alpha\beta^2 \sin^2\theta}{r(r^2 + \beta^2 \cos^2\theta)}\right)^2 + \right.$$

$$\left. + \frac{16\alpha^2 \beta^2 \sin^2\theta}{(r^2 + \beta^2 \cos^2\theta)^2}\right]^{\frac{1}{2}} - \frac{1}{2}\left[\frac{2\alpha r}{r^2 + \beta \cos^2\theta} + \frac{\alpha^2}{r^2} + \frac{2\alpha\beta^2 \sin^2\theta}{r(r^2 + \beta^2 \cos^2\theta)}\right] \qquad (3.6)$$

$$\lambda_2 = \frac{1}{2}(g_{00} + g_{33} - \sqrt{(g_{00} - g_{33})^2 + 4 g_{03} g_{30}})$$

$$+ \frac{16\alpha^2 \beta^2 \sin^2\theta}{(r^2 + \beta^2 \cos^2\theta)^2}\bigg]^{\frac{1}{2}} + \frac{1}{2}\left[\frac{2\alpha r}{r^2 + \beta \cos^2\theta} + \frac{\alpha^2}{r^2} + \frac{2\alpha\beta^2 \sin^2\theta}{r(r^2 + \beta^2 \cos^2\theta)}\right] \qquad (3.7)$$

Therefore, we take the transformation (In this case $r$ and $\theta$ are regarded as constants.)

$$dt' = \frac{-g_{00} + \lambda_2}{\lambda_2 - \lambda_1}dt - \frac{g_{30}}{\lambda_2 - \lambda_1}r\sin\theta d\varphi \qquad (3.8)$$

$$d\varphi' = \frac{1}{r\sin\theta}\frac{g_{00} - \lambda_2}{\lambda_2 - \lambda_1}dt - \frac{g_{30}}{\lambda_2 - \lambda_1}d\varphi \qquad (3.9)$$

Eq.(3.3) can be transformed into the diagonal form with

$$ds^2 = \frac{1}{2}\left\{\left[\left(2-\frac{2\alpha r}{r^2+\beta^2\cos^2\theta}+\frac{\beta^2}{r^2}+\frac{2\alpha\beta^2\sin^2\theta}{r(r^2+\beta^2\cos^2\theta)}\right)^2\right.\right.$$

$$\left.\left.+\frac{16\alpha^2\beta^2\sin^2\theta}{(r^2+\beta\cos^2\theta)^2}\right]^{\frac{1}{2}}-\frac{2\alpha r}{r^2+\beta^2\cos^2\theta}-\frac{\beta^2}{r^2}-\frac{2\alpha\beta^2\sin^2\theta}{r(r^2+\beta^2\cos^2\theta)}\right\}dt'^2$$

$$-\frac{1}{2}\left\{\left[\left(2-\frac{2\alpha r}{r^2+\beta^2\cos^2\theta}+\frac{\beta^2}{r^2}+\frac{2\alpha\beta^2\sin^2\theta}{r(r^2+\beta^2\cos^2\theta)}\right)^2+\frac{16\alpha^2\beta^2\sin^2\theta}{(r^2+\beta^2\cos^2\theta)^2}\right]\right.$$

$$-\frac{r^2+\beta^2\cos^2\theta}{r^2+\beta^2-2\alpha r}dr^2-\left(1+\frac{\beta^2}{r^2}\cos^2\theta\right)r^2d\theta^2$$

$$-\frac{1}{2}\left\{\left[\left(2-\frac{2\alpha r}{r^2+\beta^2\cos^2\theta}+\frac{\beta^2}{r^2}+\frac{2\alpha\beta^2\sin^2\theta}{r(r^2+\beta^2\cos^2\theta)}\right)^2+\frac{16\alpha^2\beta^2\sin^2\theta}{(r^2+\beta^2\cos^2\theta)^2}\right]\right.$$

$$\left.+\frac{2\alpha r}{r^2+\beta^2\cos^2\theta}+\frac{\beta^2}{r^2}+\frac{2\alpha\beta^2\sin^2\theta}{r(r^2+\beta^2\cos^2\theta)}\right\}r^2\sin^2\theta\,d\varphi'^2 \qquad (3.10)$$

The formula above has the form of Eq.(3.1), so it can be used to describe the gravitational field of static mass distribution of thin loop.

On the other hand, as we known that only by comparing with the Newtonian theory in weak gravitational fields, the solution of the Einstein's equation of gravitational field can determined, otherwise the integral constants can not be decided so that the solution is meaningless. According to this principle in the general theory of relativity, we have relation

$$g_{00}=1+2\psi \qquad (3.11)$$

Here $\psi$ is the Newtonian potential of thin loop. Now let's discuss its concrete form. As shown in Fig1, suppose the coordinates of observation point are $x_0=r'\sin\theta'\cos\phi'$, $y_0=r'\sin\theta'\sin\phi'$, $z_0=r'\cos\theta'$. The coordinates of loop at a certain point are $x=b\cos\phi$, $y=b\sin\phi$, $z=0$. The distance between these two points is

$$r=\sqrt{(x_0-x)^2+(y_0-y)^2+(z_0-z)^2}=\sqrt{r'^2+b^2-2r'b\sin\theta'\cos(\phi-\phi')} \qquad (3.12)$$

For symmetry and simplification, we can take $\phi'=0$, so the Newtonian potential of thin loop is

$$\psi=-\int\frac{Gdm}{r}=-\int_0^\pi\frac{2G\rho\,bd\phi}{\sqrt{r'^2+b^2-2r'b\sin\theta'\cos\phi}} \qquad (3.13)$$

Here $M$, $\rho$ and $b$ are the mass, line density and radius of thin loop individually. Let $\phi=\pi-\phi'$, $d\phi'=-d\phi$, $-\cos\phi=\cos\phi'=1-2\sin^2(\phi'/2)$, and put them into the formula above, we get

$$\psi = \int_{\pi}^{0} \frac{2G\rho\, bd\phi'}{\sqrt{r'^2 + b^2 + 2r'b\sin\theta - 4r'b\sin\theta \sin^2(\phi'/2)}} \tag{3.14}$$

Then let $\phi'' = \phi'/2$ again, the formula above can written as

$$\psi = -\int_{0}^{\pi/2} \frac{4G\rho\, bd\phi''}{\sqrt{r'^2 + b^2 + 2r'b\sin\theta' - 4r'b\sin\theta' \sin^2\phi''}}$$

$$= -\frac{4G\rho\, b}{\sqrt{r'^2 + b^2 + 2r'b\sin\theta'}} \int_{0}^{\pi/2} \frac{d\phi''}{\sqrt{1 - k^2 \sin^2 \phi''}} \tag{3.15}$$

In the formula, $k^2 = 4r'b\sin\theta/(r'^2 + b^2 + 2br'\sin\theta)$. Let

$$K(k^2) = \int_{0}^{\pi/2} \frac{d\phi''}{\sqrt{1 - k^2 \sin^2 \phi''}} \tag{3.16}$$

It is just the first kind of ellipse function with $k^2 < 1$ when $r' \to \infty$. So we have

$$K(k^2) = \frac{\pi}{2}\left(1 + \frac{1}{4}k^2 + \frac{9}{64}k^4 \cdots\right) = \frac{\pi}{2}\left(1 + \frac{b\sin\theta'}{r'} + \frac{9b^2 \sin^2 \theta'}{4r'^2} \cdots\right) \tag{3.17}$$

On the other hand, when $r' \gg b$, we have

$$\frac{1}{\sqrt{r'^2 + b^2 + 2r'b\sin\theta'}} = \frac{1}{r'}\left[1 - \frac{b\sin\theta'}{r'} + \frac{b^2(3\sin^2\theta' - 1)}{2r'^2} + \cdots\right] \tag{3.18}$$

Put them into Eq.(3.15), and let $2\pi\rho\, b = M$, we get

$$\psi = -\frac{GM}{r'}\left[1 - \frac{b^2(2 - 11\sin^2\theta')}{4r'^2} + \cdots\right] \tag{3.19}$$

On the other hand, we can develop $g_{\mu\nu}$ into the power series of $1/r$ and write Eq.(3.10) as

$$ds^2 = \left(1 - \frac{2\alpha}{r} + \frac{2\alpha\beta^2 \cos^2\theta}{r^3} + \cdots\right)dt'^2$$

$$-\left(1 + \frac{2\alpha}{r} - \frac{2\alpha + \beta^2 \sin^2\theta}{r^2} + \frac{12\alpha\beta^2 - 48\alpha^3}{r^3} + \cdots\right)dr^2$$

$$-\left(1 + \frac{\beta^2}{r^2}\cos^2\theta\right)r^2 d\theta^2 - \left(1 + \frac{\beta^2}{r^2} + \frac{2\alpha\beta^2 \sin^2\theta}{r^3}\cdots\right)r^2 \sin^2\theta\, d\varphi'^2 \tag{3.20}$$

When $r \to \infty$, comparing the items with $r^{-1}$ order in $g_{00}$ and $\psi$, we get according to Eq.(3.11)

$$1 - \frac{2\alpha}{r} = 1 - \frac{2GM}{r'} \tag{3.21}$$

Let $\alpha = GM$, we get $r = r'$. However, the formula above is only the corresponding relation when the mass is concentrated at the center point of thin loop. It is not the gravitational potential of thin loop. In order to obtain the potential of thin loop, we should consider higher order items. There are no items with $r^{-2}$ order in both Eqs.(3.19) and (3.20). By considering the items containing $r^{-3}$ order, we have

$$1 - \frac{2GM}{r} + \frac{2GM\beta^2 \cos^2\theta}{r^3} = 1 - \frac{2GM}{r'} + \frac{2GMb^2(0.5 - 2.75\sin^2\theta')}{r'^3} \quad (3.22)$$

Thus we see that the function forms are different if high order items are considered. It means that the Einstein's theory of gravity can not asymptotically coincide with the Newtonian theory automatically in general. In order to let them asymptotically coinciding to each other, further transformation is needed. Because constant $\beta$ has the dimension of length, in the problem of thin loop, we can let $\beta = b$. Because we always have $\cos^2\theta \geq 0$, but may have $0.5 - 2.75\sin^2\theta' < 0$ in some cases, so in general we have $0.5 - 2.75\sin^2\theta' \neq \cos^2\theta$. But we can let $\theta = \theta'$. In this way, Eq.(3.22) becomes

$$\frac{1}{r} - \frac{b^2 \cos^2\theta'}{r^3} = \frac{1}{r'} - \frac{b^2(0.5 - 2.75\sin^2\theta')}{r'^3} \quad (3.23)$$

Let $A = b^2\cos^2\theta'$, $B = b^2(0.5 - 2.75\sin^2\theta')$. Because of $r' \gg b$, the only real solution of the third order equation above is

$$\frac{1}{r} = \left[\frac{1}{2Ar'^3}(B - r'^2) + \frac{i}{2A^{3/2}r'^3}\sqrt{4r'^6 - A(B - r'^2)^2}\right]^{\frac{1}{3}}$$

$$+ \left[\frac{1}{2Ar'^3}(B - r'^2) - \frac{i}{2A^{3/2}r'^3}\sqrt{4r'^6 - A(B - r'^2)^2}\right]^{\frac{1}{3}}$$

$$= (a + ib)^{1/3} + (a - ib)^{1/3} = 2Q\cos(\delta/3) \quad (3.24)$$

Here $a = (B - r'^2)/(2Ar'^3)$, $b = \sqrt{4r'^6 - A(B - r'^2)^2}/(2A^{3/2}r'^3)$, $Q = (a^2 + b^2)^{1/6}$, $\delta = tg(b/a)$.
In this way, we can write $r = r(r', \theta')$ and obtain

$$dr = \frac{dr}{dr'}dr' + \frac{dr}{d\theta'}d\theta' = T(r', \theta')dr' + V(r', \theta')d\theta' \quad (3.25)$$

The concrete forms of functions $T(r', \theta')$ and $V(r', \theta')$ are unimportant, so the concrete forms are not written out here. Now put the relation into Eq.(3.10), the solution of gravitational equation of thin loop is reached with the form of Eq.(3.2)

$$ds^2 = \frac{1}{2}\left\{\left[\left(2 - \frac{2\alpha r}{r^2 + b^2\cos^2\theta'} + \frac{b^2}{r^2} + \frac{2\alpha b^2 \sin^2\theta'}{r(r^2 + b^2\cos^2\theta')}\right)^2\right.\right.$$

$$\left.+ \frac{16\alpha^2 b^2 \sin^2\theta'}{(r^2 + b\cos^2\theta')^2}\right]^{\frac{1}{2}} - \frac{2\alpha r}{r^2 + b^2\cos^2\theta} - \frac{b^2}{r^2} - \frac{2\alpha b^2 \sin^2\theta'}{r(r^2 + b^2\cos^2\theta')}\right\}dt'^2$$

$$- \frac{r^2 + b^2\cos^2\theta'}{r^2 + b^2 - 2\alpha r}T^2(r', \theta')dr'^2 - \left[\left(1 + \frac{b^2}{r^2}\cos^2\theta'\right)\frac{r^2}{r'^2}\right.$$

$$\left.+ \frac{r^2 + b\cos^2\theta'}{r^2 + b^2 - 2\alpha b}\frac{V^2(r', \theta')}{r'^2}\right]r'^2 d\theta'^2 - \frac{r^2}{2r'^2}\left\{\left[\left(2 - \frac{2\alpha r}{r^2 + b^2\cos^2\theta'}\right.\right.\right.$$

$$+\frac{b^2}{r^2}+\frac{2ab^2 r\sin^2\theta'}{r(r^2+b^2\cos^2\theta')^2}\Bigg)^2+\frac{16\alpha^2 b^2\sin^2\theta'}{(r^2+b^2\cos^2\theta')^2}\Bigg]^{\frac{1}{2}}$$

$$+\frac{2\alpha\ r}{r^2+b^2\cos^2\theta'}+\frac{b^2}{r^2}+\frac{2\alpha b^2\sin^2\theta'}{r(r^2+b^2\cos^2\theta')}\Bigg\}r'^2\sin^2\theta'\ d\varphi'^2$$

$$-\Bigg[2\frac{r^2+b^2\cos^2\theta'}{r^2+b^2-2\alpha\ r}\frac{T(r',\theta')V(r',\theta')}{r'}\Bigg]r'dr'd\theta' \qquad (3.26)$$

Here $r=r(r',\theta')$. Because Eqs. (3.24) and (3.26) are too complex, we discuss the gravitational field of thin loop from relation (3.23) directly. It is known that when $r'=0$ we have $r=0$. So when $r'=0$, we have $g_{00}\to\infty$, $g_{22}\to\infty$, and $g_{33}\to\infty$. The results show that a singularity appears in the centre point of thin loop. This singularity is exposed in vacuum completely, no matter how much the mass of thin loop is, big or small. In the nearby region of the center point, space is also high curved. This result is absolutely absurd for it obviously violate common knowledge. It does not like the Schwarzschild solution in which the singularity is hided in the center of big mass so that it can not be observed directly. In this case, it seems that physicists can bear the existence of singularity. But we can not bear the singularity to appear in the centre point of a thin loop with a small mass without being covered.

Besides, space nearby thin loop's surface is also high curved. Take $\alpha\to 0$ and $\theta'=\pi/2$ on the surface of thin loop. In this case, Eq.(3.26) becomes

$$ds^2 = dt'^2 - \frac{r^2}{r^2+b^2}T^2(r',\theta')dr'^2 - \Bigg[\frac{r^2}{r'^2}+\frac{r^2V^2(r',\theta')}{r'^2(r^2+b^2)}\Bigg]r'^2 d\theta'^2$$

$$-\Bigg(\frac{r^2}{r'^2}+\frac{b^2}{r'^2}\Bigg)r'^2\sin^2\theta'\ d\varphi'^2 - \frac{2r^2}{r'(r^2+b^2)}T(r',\theta')V(r',\theta')r'dr'd\theta' \qquad (3.27)$$

Let $b=0.67$, Eq.(3.23) becomes

$$\frac{1}{r}=\frac{1}{r'}+\frac{1}{r'^3} \qquad \text{or} \qquad r=\frac{r'^3}{r'^2+1} \qquad (3.28)$$

$$T(r',\theta')=\frac{r'^3+3r'^2}{(r'^2+1)^2} \qquad V(r',\theta')=0 \qquad (3.29)$$

Taking $r'=b=0.67$, we have $r=0.21$, $T=1$. Put them into Eq.(3.27), we obtain $g'_{11}=-0.09$, $g'_{22}=-0.01$, $g'_{33}=-1.01$. It is obvious that the space nearby the surface of thin loop is high curved. This does not agree with practical situation completely. On the surface of thin loop, for such a weak gravitational field, the space should be nearly flat with $g'_{11}=g'_{22}=g'_{33}\approx -1$. Because these are measurable quantities, it can be said that the Einstein's theory of gravity is unsuitable for the problem of mass distribution of thin loop.

On the other hand, let $r^2+b^2-2\alpha\ r=0$ in Eq.(3.26), we have $r=\alpha\pm\sqrt{\alpha^2-b^2}$. By taking $M=1Kg$ for weak field, we have $\alpha\sim GM/c^2=7.41\times 10^{-28} m$. Because of $\alpha\ll b$, $r$ is not a real number in this case. The second singularity decided by the relation $r^2+b^2-2\alpha\ r=0$ in Eq.(3.26) does not exist in general. But according to the Newtonian theory, according to Eq.(3.13) at the loop's center point $r' = 0$, gravitational potential is a limit constant with

$$\psi = -\int_0^\pi 2G\rho \, d\phi = -2\pi \, G\rho = -\frac{GM}{b} \quad (3.30)$$

So gravity at the center point is zero. This agrees with practical situation. Therefore, it should be noted that in order to decide the integral constants of the solutions of the Einstein's equation of gravitational fields, we have to let the Einstein's theory asymptotically coinciding with the Newtonian theory under the condition of weak fields with $r \to \infty$. Otherwise we can not compare the Einstein's theory with the Newtonian one, so that we can not determinate the validity of the Einstein's gravity theory. But in the problem of static axial symmetry distribution of thin loop mass, singularity appears at the center point of loop in vacuum so that the theory becomes meaningless. On the region nearby the center and surface of loop, space-time is high curved. All of these do not agree with practical situations.

The situation when the cross section of thin loop is not zero is discussed below. In this case, the gravitation field is with three parameters. The third is the radium of loop's cross section. On the other hand, as we known that the Kerr-Newman metric is one with axial symmetry and three parameters [11]. At present, it is used to describe the external gravitational field of charged revolving sphere. If the solution of the Einstein's equation of gravitational field with three parameters and axial symmetry is unique, through the coordinate transformation from the Kerr-Newman metric, we can also reach the gravitational field of loop with cross section. By the same method of diagonalization, we can write the Kerr-Newman metric as

$$ds^2 = \frac{1}{2}\left\{\left[\left(2 - \frac{2\alpha r - Q^2}{r^2 + \beta^2 \cos^2\theta} + \frac{\beta^2}{r^2} + \frac{2\alpha\beta^2 \sin^2\theta}{r(r^2 + \beta^2 \cos^2\theta)}\right)^2\right.\right.$$

$$\left.\left. + \frac{16\alpha^2\beta^2 \sin^2\theta}{(r^2 + \beta^2 \cos^2\theta)^2}\right]^{\frac{1}{2}} - \frac{2\alpha r - Q^2}{r^2 + \beta^2 \cos^2\theta} - \frac{\beta^2}{r^2} - \frac{2\alpha\beta^2 \sin^2\theta}{r(r^2 + \beta^2 \cos^2\theta)}\right\}dt'^2$$

$$- \frac{r^2 + \beta^2 \cos^2\theta}{r^2 + \beta^2 - 2\alpha r - Q^2}dr^2 - \left(1 + \frac{\beta^2}{r^2}\cos^2\theta\right)r^2 d\theta^2$$

$$-\frac{1}{2}\left\{\left[\left(2 - \frac{2\alpha r - Q^2}{r^2 + \beta^2 \cos^2\theta} + \frac{\beta^2}{r^2} + \frac{2\alpha\beta^2 \sin^2\theta}{r(r^2 + \beta^2 \cos^2\theta)}\right)^2 + \frac{16\alpha^2\beta^2 \sin^2\theta}{(r^2 + \beta^2 \cos^2\theta)^2}\right]^{\frac{1}{2}}\right.$$

$$\left. + \frac{2\alpha r - Q^2}{r^2 + \beta^2 \cos^2\theta} + \frac{\beta^2}{r^2} + \frac{2\alpha\beta^2 \sin^2\theta}{r(r^2 + \beta^2 \cos^2\theta)}\right\}r^2 \sin^2\theta \, d\varphi'^2 \quad (3.31)$$

Here constant $Q$ is relative to the charge of sphere. When $r \gg \alpha$ and $r \gg \beta$, we can get from formula above

$$g_{00} = 1 - \frac{2\alpha}{r} + \frac{Q^2}{r^2} + \frac{2\alpha\beta^2 \cos^2\theta}{r^3}^2 + \cdots \quad (3.32)$$

On the other hand, when the area of thin loop's cross section is considered, the function form of the Newtonian potential is very complex. We can get the same conclusion by a simple estimate without accurate calculation. Suppose the radium of thin loop's cross section is $h$, when $r \gg b$, $r \gg h$ and $h \sim b$, we can always write the Newtonian gravitational potential of thin loop as

$$\psi = -\frac{GM}{r'}\left[1 + \frac{f_1(\theta',b,h)}{r'} + \frac{f_2(\theta',b,h)}{r'^2} + \cdots\right] \tag{3.33}$$

From discussion above, we can take $\theta = \theta'$, $\beta = b$. When $r \to \infty$, we can only remain the item containing $r^{-2}$ and get

$$-\frac{GM}{r} + \frac{Q^2}{2r^2} = -\frac{GM}{r'} + \frac{f_1(\theta',b,h)}{r'^2} \tag{3.34}$$

Let $x = 1/r$, $x' = 1/r'$, we have from formula above

$$x = \frac{GM + \sqrt{(GM)^2 + 2Q^2(f_1 x'^2 - GMx')}}{Q^2} \tag{3.35}$$

Put it into Eq.(3.31), we can get the metric of loop. It is easy to know from the formulas that when $r' = 0$ ($x' = \infty$), we have $r = 0$ ($x = \infty$). So it can also be seen from Eq.(3.33) that there exists still singularity at the center point of loop, and the singularity is also exposed in vacuum completely. Space nearby the center point and the surface of loop is also high curved. The situation is completely the same as that when the area of cross section of thin loop is not taken into account.

## 3. Double sphere's static gravitational field with axial symmetry

The static gravitational field of mass double sphere is discussed below. As shown in Fig. 3.2, the masses and radius of both spheres are $M$ and $b$ individually. The centers of two spheres are at the points $\pm b$ of the $z$ axis. The gravitational field of this axial symmetrical distribution with two parameters can also be obtained through the coordinate transformation of the Kerr solution. For this problem, the Newtonian potential of gravity is

$$\psi = -GM\left(\frac{1}{r_1} + \frac{1}{r_2}\right) = -GM\left(\frac{1}{\sqrt{r'^2 + b^2 + 2br'\cos\theta}} + \frac{1}{\sqrt{r'^2 + b^2 - 2br'\cos\theta}}\right) \tag{3.36}$$

When $r' \gg b$, we have

$$\psi = -\frac{2GM}{r'}\left(1 - \frac{b^2 - 3b^2\cos^2\theta}{2r'^2}\right) \tag{3.37}$$

From Eq.(3.11), we get relation

$$1 - \frac{2\alpha}{r} + \frac{2\alpha\beta^2\cos^2\theta}{r^3} = 1 - \frac{4GM}{r'} + \frac{2GMb^2(1 - 3\cos^2\theta')}{r'^3} \tag{3.38}$$

Let $\alpha = GM$, $\beta = b$ and $\theta = \theta'$, we have

$$\frac{1}{r} - \frac{b^2\cos^2\theta'}{r^3} = \frac{1}{r'} - \frac{b^2(1 - 3\cos^2\theta')}{2r'^3} \tag{3.39}$$

Let $A = b^2\cos^2\theta'$, $B = b^2(1 - 3\cos^2\theta')/2$, from Eq.(3.39), we can also obtain the formula similar to Eq.(3.24). Put it into Eq.(3.10), the metrics of mass double sphere distribution is also be obtained. For simplification, we can discuss directly from Eq.(3.39). When $r' = 0$, we have $r = 0$ so that $g_{00} \to \infty$, $g_{22} \to \infty$, $g_{33} \to \infty$ as well as $g_{23} \to \infty$ ($T \neq 0$, $V \neq 0$). The singularity appears at the point at which two spheres contact each other. Suppose $b = \sqrt{2}$ and $\theta = \pi/2$, Eq.(3.39) becomes

$$\frac{1}{r} = \frac{1}{r'} - \frac{1}{r'^3} \qquad \text{or} \qquad r = \frac{r'^3}{r'^2 - 1} \qquad (3.40)$$

Let $M = 1 Kg$, i.e., the gravitational field is very weak so that we can let $\alpha \to 0$ in Eq.(3.26). The formula similar to Eq.(329) can be obtained with

$$T(r', \theta') = \frac{r'^4 - 3r'^2}{(r'^2 - 1)^2} \qquad V(r', \theta') = 0 \qquad (3.41)$$

Let $r' = 2$ at some points on the surface of double spheres, we get $r = 2.67$ and $T = 0.44$. Put them into Eq.(3.27), we obtain $g_{11} = -0.34$, $g_{22} = -1.78$ and $g_{33} = -2.28$. It means that the space nearby the surfaces of two spheres is also high curved. It is also obvious that this result does not agree with practical situation completely. For this weak field, space should be nearly flat with $g_{11} = g_{22} = g_{33} \approx -1$. More serious problem is that when $r' < 1$, according to Eq.(3.40), $r$ becomes a negative number so that it is meaningless. So it can be said that the Einstein's theory of gravity is unsuitable for the problem of double spherical mass distribution.

In fact, there are many other axial symmetry distributions of static masses with double and three parameters. For example, three spheres are superposed one by one in a line, two cones are superposed with their cusps meeting together and hollow column and so do. All of their gravitational fields should be obtained by means of the coordinate transformations of the Kerr and Kerr-Newman solutions in principle. But as shown above, the same problems would be caused.

## 4. Discussion for general situations

Therefore, we can conclude from discussions above

1. If the axial symmetry solutions of the Einstein's equations of gravitational fields with double and three parameters are unique, the gravitational fields of the static mass axial symmetry distributions of thin loop and double spheres should be obtained by the coordinate transformations of the Kerr and Kerr-Newman solutions. However, these solutions can not coincide with practical situations. Therefore, the Einstein's theory of gravity can not be a universally suitable one. If the axial symmetry solutions of the Einstein's equations of gravitational fields with double and three parameters are not unique, i.e., there exist other solutions for the same Einstein's equation which can describe these problems well (even though they have not be founded now), the uniqueness which is a basic demand for a universal physical theory would be destroyed.

2. A great number of theories about space-time singularity, black holes, white holes and wormholes have established based on the general theory of relativity. In common viewpoint, these objects with space-time singularity are caused by the distributions of high density and huge masses. However, from discussions above, even for the systems of thin loop and two small spheres, singularities would also appear based on the Einstein's theory of gravity. This fact shows that singularities are not caused actually by the high density and huge masses. They only exist in the Einstein's theory actually, not in nature. The singularity has nothing to do with real world. So-called black holes, white holes and wormholes with space-time singularity are actually illusive objects. This problem will be further discussed later.

3. As well-known that only by comparing with the Newtonian theory under the condition of weak

fields, the solution forms of the Einstein's equations of gravitational fields can be finally determined. According to the present method, under the condition of weak field, let $h_{\mu\nu}$ to be a small quantity with

$$g_{\mu\nu} = \eta_{\mu\nu} + h_{\mu\nu} \tag{3.42}$$

Here $\eta_{\mu\nu}$ is the Minkowski metric. In this case, the Einstein's equations become [12]

$$\Box^2 h_{\mu\nu} - \frac{\partial^2}{\partial x^\lambda \partial x^\mu} h_\nu^\lambda - \frac{\partial^2}{\partial x^\lambda \partial x^\nu} h_\mu^\lambda + \frac{\partial^2}{\partial x^\mu \partial x^\nu} h_\lambda^\lambda = -16\pi G S_{\mu\nu} \tag{3.43}$$

$$S_{\mu\nu} = T_{\mu\nu} - \frac{1}{2} \eta_{\mu\nu} T_\lambda^\lambda \tag{3.44}$$

By introducing proper function $\varepsilon^\mu$ and taking coordinate transformations below

$$x^\mu \to x^\mu + \varepsilon^\mu(x^\mu) \qquad h_{\mu\nu} \to h_{\mu\nu} - \frac{\partial \varepsilon_\mu}{\partial x^\nu} - \frac{\partial \varepsilon_\nu}{\partial x^\mu} \tag{3.45}$$

the harmonious coordinate condition $g^{\mu\nu}\Gamma^\lambda_{\mu\nu} = 0$ can be satisfied, so that we have

$$\frac{\partial}{\partial x^\mu} h_\nu^\mu = \frac{1}{2} \frac{\partial}{\partial x^\nu} h_\mu^\mu \tag{3.46}$$

The energy and momentum tensors shown in Eq.(3.43) should also be transformed correspondingly. Put the relations into Eq.(3.43), the equation becomes

$$\Box^2 h_{\mu\nu} = -16\pi\, G S_{\mu\nu} \tag{3.47}$$

The equations has the solution similar to classical retarded potential with

$$h_{\mu\nu}(\vec{x},t) = \int \frac{4 G S_{\mu\nu}(\vec{x}', t-r)}{r} d^3 \vec{x}' \tag{3.48}$$

Here $r = |\vec{x} - \vec{x}'|$, $h_{00} = 2\psi$, $\psi$ is the Newtonian potential. In this way, it seems that the Einstein's theory is proved to coincide with the Newtonian theory under the condition of weak field. However, by careful examination, we would find that the thing is not so simple. In order to reach Eq.(3.48), we have to introduced coordinate transformation (3.45). Meanwhile, the same transformation should be taken for Eq.(3.44). It means that the mass and energy's distribution form of original system would be changed and the original solution would be transformed into other one with different symmetry. In this way, the new solution (3.48) is meaningless for the original problem we wand to discuss. This problem will be discussed again in next paper, for it involves the rationality problem of the principle of general relativity. Here we only provide an example to show this conclusion. That is the so-called Kasner metric [13] for the solution of the Einstein's equation with infinite mass line (or column) distribution

$$ds^2 = r^{2a} dt^2 - dr^2 - r^{2b} d\varphi^2 - r^{2c} dz^2 \tag{3.49}$$

In general, we have $a \neq 0$, $b \neq 0$ and $c \neq 0$, otherwise the solution becomes the Minkowski metric of flat space-time. According to the Newtonian theory, for the same problem, the strength of gravitational field external line (or column) is

$$E = -\frac{G\rho}{r} \tag{3.50}$$

Here $\rho =$ constant is line mass density. When $r$ is big enough, $E$ is small enough and the field can be considered as a weak one. In this case we have $g_{00} = r^{2a} = 1 + 2\psi$, or

$$\psi = \frac{1}{2}\left(r^{2a} - 1\right) \tag{3.51}$$

So when $r$ is big enough, from formula above, we get

$$E = -\frac{d\psi}{dr} = -ar^{2a-1} \tag{3.52}$$

Comparing with the result of Newtonian theory according to Eq.(3.50), we should have

$$\frac{G\rho}{r} = ar^{2a-1} \tag{3.53}$$

However it is obvious that when $a \neq 0$, the function forms on the two sides of equation are completely different so that it is impossible to compare them. When $a = 0$, the right side of the formula is equal to zero but the left side is not, so the equation does not exist. In fact, when $r \to \infty$, we have

$$\lim_{r \to \infty} r^{2a} = \begin{cases} 0 & a > 0 \\ \infty & a < 0 \end{cases} \quad \lim_{r \to \infty} r^{2b} = \begin{cases} 0 & b > 0 \\ \infty & b < 0 \end{cases} \quad \lim_{r \to \infty} r^{2c} = \begin{cases} 0 & c > 0 \\ \infty & c < 0 \end{cases} \tag{3.54}$$

It is obvious that all metrics $g_{00}$, $g_{22}$ and $g_{33}$ in Eq.(4.49) can not be written as the forms of Eq.(3.42), so that we can not connect the Einstein's theory with the Newtonian one, i.e., the Einstein's theory can not coincide asymptotically with the Newtonian theory. Because we can not establish relations between constants $G$, $\rho$ and $a$, $b$ and $c$ so that the constants $a$, $b$ and $c$ can not be determined, the solution (3.49) is meaningless actually for the problem of static mass line distribution. In fact, the Einstein's theory and the Newtonian theory are two completely different systems with completely different starting points. It is impossible for them to reach asymptotical consistent under the conditions of weak fields in general.

The common procedure to obtain the solutions of the Einstein's equation of gravity is firstly to simplify the metrics based on some space-time symmetry, then to solve the equation. Up to now, a lot of solutions have been founded. But many of them are considered meaningless in physics for no practical systems can be founded corresponding to them. However, the real situations may be that

1. Under the conditions of weak fields, the solutions of the Einstein's theory can not asymptotical coincide with the Newtonian theory. In this case, the integral constants in the solutions of the Einstein's equations can not be determined so that we can say that the Einstein's theory of gravity is actually unsuitable for these problems.

2. Though the solutions of the Einstein's equations of gravitational fields can coincide with the Newtonian theory in weak fields, but they are obviously irrational in general situations.

At present, if the results of the Newtonian theory and the Einstein's theory are different, we always consider that the Newtonian theory is wrong. But should we ask whether the Einstein's theory of gravity is alright? For weak gravitational fields, the Newtonian theory goes through so many verifications and can be considered basically correct. For the same problems under the condition of weak fields, if the Einstein's theory can not asymptotically coincide with the Newtonian theory, how can we always say that the Einstein's theory is wrong? Besides suspecting the correctness of the Einstein's theory, we have no other outlet. It is not a scientific attitude when we find that the Einstein's theory can not asymptotically coincide with the Newtonian theory for a certain problem, we only say that this solution of Einstein's equation is meaningless in physics then sent it away randomly. We can not help to ask that if these solutions are meaningless, where are the meaningful ones for the same problems?

Up to new, only the Schwarzschild solution obtained four verifications actually. Speaking strictly, the solution was proved effective only in the weak gravitational field of the sun. The verifications are too littler comparing with the Newtonian theory of gravity, quantum mechanics and special relativity. On the hand, under the condition of strong field, space-time singularity appears in the Schwarzschild solution. So it may be said that the correctness of the Schwarzschild solution is only a coincidence. It is unsuitable to regard the Einstein's theory of gravity as a foundational theory of interaction. As we know that though we need lots of proofs to verify a theory, only a proof can overthrow a theory sometimes. Physicists should keep their brains clear for the Einstein's theory of gravity. It is unadvisable for physicists to lose their judgment ability only by the great authority of Einstein and the beautiful form of the theory. Physics is an experimental science to pursue reality. The beauty of format is not main aim. Besides the Einstein's theory, there are many other theories of gravity now. But most of them are established based on the concept of curved space-time. Therefore, all of them are facing the same problems which exist in the Einstein's theory. Owing to the problems mentioned in the paper, we should survey the rationality of the Einstein's theory.

Because in the weak gravitational field of the sun, the Einstein's theory of gravity is the most simple and effective one, it would have some rationality. As shown in next paper, the spherical symmetry solution of the Einstein's equation of gravity is transformed into flat space-time to describe. The results show that the experiments to support the general theory of relativity can also be explained. But the theory has no any singularity in strong field. Besides, there are more experiments and astronomic observations can be rationally explained. Based on this result, a more rational theory of gravity can be established in the form of electromagnetic theory with the Lorentz invariability. In this way, the gravitational and electromagnetic interactions can be described in a consistent form. Similar to electromagnetic theory, this kind of theory of gravity is easy to quantization and normalization. The detail will be provided in next paper.